\documentclass[epj]{svjour}
\usepackage{graphicx}
\usepackage{rotating}
\graphicspath{{Images/}}

\begin{document}
\title{Hydrogen targets for exotic-nuclei studies developed over the past 10 years}
\author{A.~Obertelli \inst{1} \and T.~Uesaka \inst{2}}
\institute{CEA, Centre de Saclay, IRFU/Service de Physique Nucl\'eaire, F-91191 Gif-sur-Yvette, France \and RIKEN Nishina Center, Saitama 351-0198, Japan}

\date{Received: date/ Revised version: date}

\abstract{
Hydrogen-induced reactions provide essential information on nuclear structure, complementary to other experimental probes.
For studies at both low and relativistic incident energy, developments in hydrogen targets have been performed over the past 10 years in parallel with the development of new radioactive beams. We present a review of all major hydrogen target developments related to the study of exotic nuclei with direct reactions in inverse kinematics. Both polarized and non-polarized systems are presented.
\PACS{29.25.Pj}
}

\maketitle  

\section{Introduction}
\begin{figure}
\begin{center}
\includegraphics[trim= 0mm 0mm 40mm 90mm,clip,width=7cm]{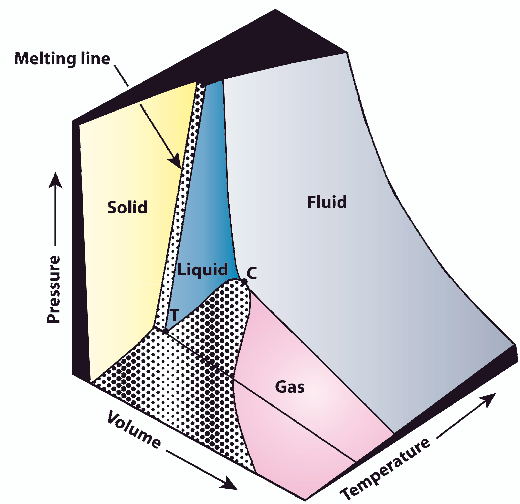}
\caption{Pressure-volume-temperature phase diagram for pure hydrogen. The triple point T lies at T=14K and P=70 mbars.}
\label{fig1}
\end{center}
\end{figure}

\begin{figure}[t]
\begin{center}
\includegraphics[trim= 0mm 0mm 0mm 0mm,clip,width=8cm]{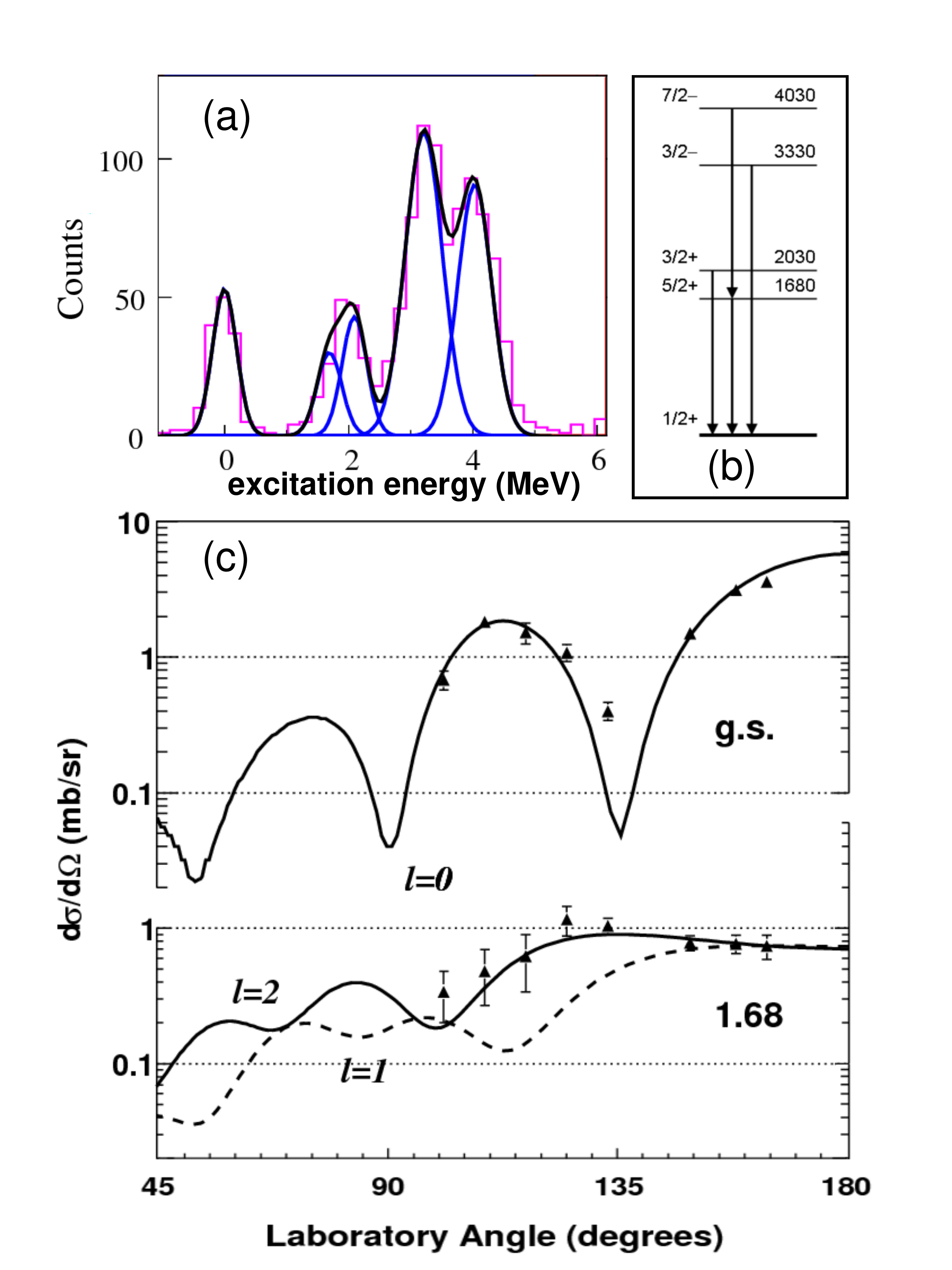}
\caption{(Color online) (a) Excitation-energy spectrum of $^{25}$Ne obtained from the one-nucleon transfer measurement $^{24}$Ne(d,p)$^{25}$Ne at 10 MeV/nucleon in inverse kinematics~\cite{PhysRevLett.104.192501}. (b) Level scheme obtained from triple particle-gamma-residue coincidences. (c) Angular distributions in coincidence with the ground state and the first excited state at 1.68 MeV in $^{25}$Ne.  The sensitivity to the transfered angular momentum $\ell$ is obtained from the angular distributions for a given populated state. Courtesy W. N. Catford (University of Surrey).}
\label{fig2}
\end{center}
\end{figure}

\begin{figure}[t]
\begin{center}
\includegraphics[trim= 30mm 0mm 40mm 0mm,clip,width=7.5cm]{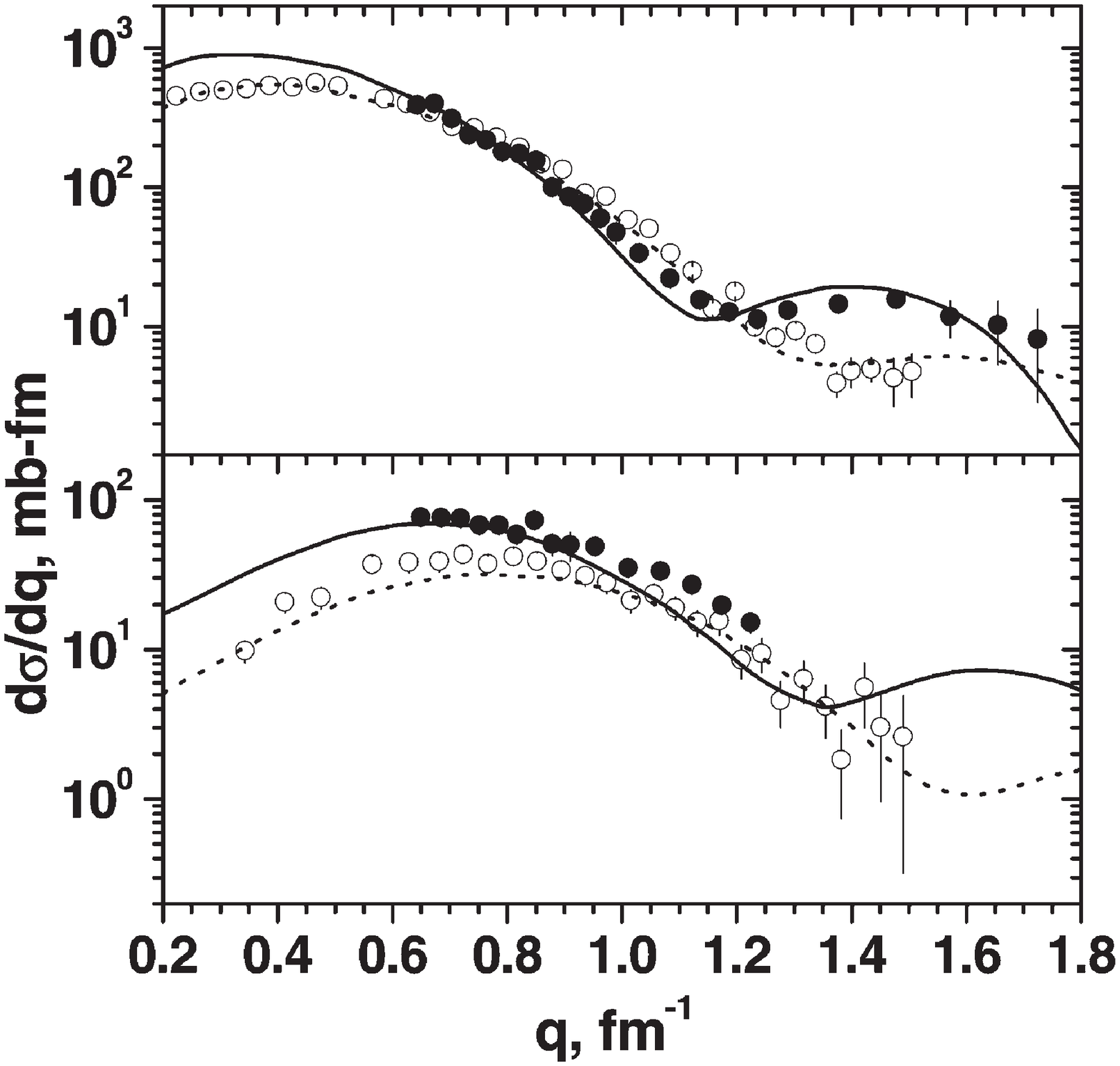}
\caption{The elastic (top) and inelastic (bottom) cross sections as a function of the transfered momentum. The data taken with 24.5 A MeV and 40.9A MeV beams are shown by the solid and open circles, respectively. Lines result from calculations assuming a neutron "halo" in $^6$He. Continuous and dashed  lines correspond to an incident energy of 24.5 A MeV and 40.9 A MeV, respectively. Reprinted from~\cite{Stepantsov200235}, with permission from Elsevier. }
\label{fig3}
\end{center}
\end{figure}

\begin{figure}[t]
\begin{center}
\includegraphics[trim= 0mm 0mm 0mm 50mm,clip,width=7.5cm]{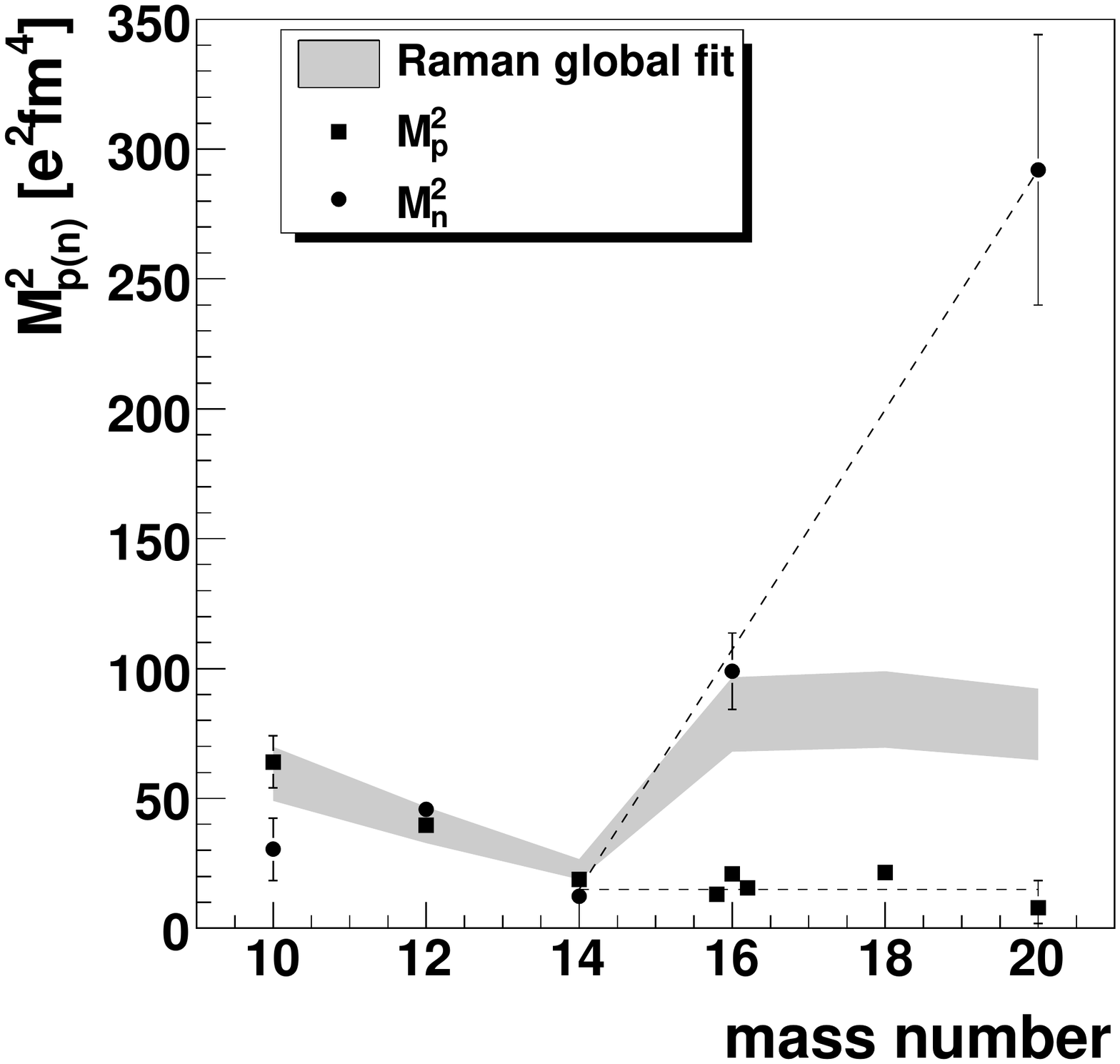}
\caption{Comparison of the measured electromagnetic (Mp$^2$) and neutron (Mn$^2$) transition probabilities for even-even carbon isotopes~\cite{PhysRevC.79.011302}. Courtesy Z. Dombradi (ATOMKI laboratory).}
\label{fig4}
\end{center}
\end{figure}

Hydrogen-induced reactions have strong advantages in nuclear structure studies, as it has been extensively shown in direct kinematics experiments by use of proton and deuteron beams on stable-nucleus targets.  Hydrogen targets allow to take benefit from these advantages with exotic beams, in inverse-kinematics reactions. There are at least four advantages: (i) simplicity, (ii) specificity, (iii) luminosity and (iv) capability of polarization, which are enhanced in some cases by introducing pure hydrogen targets, despite the technical challenge of a solid or liquid phase at very low temperatures (see Fig.~\ref{fig1}).\\
The very first argument in favor of hydrogen-induced reactions  is that protons are structureless at energies involved in nuclear-structure studies with incident energies usually lower than 1 GeV/nucleon. The $\Delta$ resonance of the nucleon, its first excited state, lies at an excitation energy of 290 MeV. Therefore, in proton-induced reactions in inverse kinematics, the only unknown structure comes from the projectile to be studied. When deuteron is used, its wave function and the contribution to the breakup channel have to be considered in the analysis. These last points are considered under controle and have been validated on several data sets.\\
 The target material itself, hydrogen, offers unique capabilities in terms of sensitivity of nuclear single-particle structure and selectivity. It is used to perform inelastic excitations, nucleon pickup such as $(d,p)$ reactions and proton induced knockout reactions such as $(p,2p)$, $(p,pn)$ and possibly nucleon-pair removal $(p,3p)$ which has not been much investigated in exotic nuclei, partly due to the low cross sections. Each of these reaction mechanisms highlights specific structure aspects of the nucleus. \\
Low-energy reactions such as neutron pickup $(d,p)$ and stripping $(p,d)$ offer a unique way to probe the shell structure of nuclei. With a sufficient statistics (requesting about 10$^4$ pps of beam), angular distributions related to the population of a given state allow an unambiguous assigment of the transfered angular momentum $\ell$ and the cross section gives access to the shell occupancy through spectroscopic factors. An illustration is given in Fig.~\ref{fig2} obtained from the one-neutron pickup $^{24}$Ne(d,p$\gamma$)$^{25}$Ne at 10 MeV/nucleon~\cite{PhysRevLett.104.192501} at the SPIRAL facility of the GANIL laboratory. The measurement of the protons' scattering angle and energy, when produced in the binary reaction $(d,p)$, leads to the reconstruction of the excitation-energy spectrum of $^{25}$Ne for both bound and unbound states. The target thickness strongly impacts the energy resolution since the reaction vertex position inside the target is not known. The angular distribution corresponding to the population of a given final state of $^{25}$Ne allows the extraction of the transfered angular momentum which constrains the spin assignment for each final state.\\
One proton removal $(p,2p)$ reaction at relativistic energies of several hundreds of MeV probes the single-particle structure of nuclei below the Fermi surface. The accessible information has been shown to be consistent with results from $(e,e'p)$ measurements and (d,$^3$He) proton pickup at lower incident energy~\cite{kramer}. The proton-induced knockout brings more distorsions than the $(e,e'p)$ technique since the nuclear interaction is less known and effects from initial and final state nuclear interactions have to be deconvoluted to obtain nuclear-structure information. Nevertheless to study exotic nuclei, only the $(p,2p)$ can be used since no electron-exotic beam collider exists yet. On the same line, the neutron density distribution of a nucleus can be probed via $(p,pn)$ reactions. Ideally, such experiments have to be performed in a rather narrow incident-energy range from about 200 MeV/nucleon to 500 MeV/nucleon in order to stay in the direct-reaction regime and minimize both entrance and exit channel final-state interactions.\\
Proton-induced inelastic scattering is sensitive to the nuclear radius and the nuclear collectivity of a given state. The measurement of differential cross section and their analysis by use of optical potentials and the coupled-reaction channel (CRC) formalism, or distorted-wave Born approximation (DWBA) as a first analysis, can end up to a refined understanding of the nuclear structure of the populated states~\cite{Lapoux200118}. An example is given in Fig.~\ref{fig3}: the elastic and inelastic cross section to the first excited state of $^{6}$He performed at  24.5 A MeV and 40.9 A MeV allowed a detailed investigation of the halo structure of $^6$He~\cite{Stepantsov200235}. These measurements were performed with CH$_2$ targets of about 1 mg/cm$^{2}$ at intensities of 10$^6$ pps. Proton-induced inelastic scattering is complementary to Coulomb excitation (or lifetime measurement) to understand the nature of collectivity of specific low-lying states. Inelastic excitation by use of a target sensitive to the mass quadrupole moment allows to separate the neutron and proton contributions of a specific excitation,  when combined with Coulomb excitation. Proton inelastic scattering is 2-3 times more sensitive to the neutron distribution than to the charge distribution examined by use of the Coulomb excitation. In case the beam intensity does not allow the measurement of differential cross sections, inclusive inelastic scattering measurements can still provide information on the collectivity of a nucleus. The resolving power of the method is illustrated in Fig.~\ref{fig4} for the case of the population of the first excited 2$^+$ state in heavy C isotopes, where the B(E2) values obtained experimentally (M$_p^2$) and deduced from the systematics are compared with the neutron transition probabilities (M$_n^2$)~\cite{PhysRevC.79.011302}.\\

Using pure hydrogen as a target is a way to maximize the number of scattering centers while conserving a satisfying energy resolution. Iin most of the inverse-kinematics reactions, the target thickness is determined by a compromise between the luminosity and the energy resolution. Indeed, the energy resolution (either from particle detection of gamma detection) comes most of the time from energy loss, straggling or re-interaction in the target.\\ 
A strong argument in favor of a pure hydrogen target compared to composite targets, such as polypropylene CH$_2$, is related to the background in the final spectrum, either for particle detection or $\gamma$ spectroscopy, depending on the studied reaction. Pure hydrogen targets have the considerable advantage to avoid contamination from carbon-induced reactions. In the case of an inclusive inelastic cross section measurement, one needs to perform, in addition, a measurement on a carbon target in order to substract the carbon contribution. This costs nearly twice the beam time to extract proper hydrogen-induced cross sections. Such procedure, besides being beam-time consuming, generally leads to extra uncertainties.\\

Spin polarization enhances potential of research with hydrogen-induced reactions. From recent experimental and theoretical studies of unstable nuclei, it is obvious that spin degrees of freedom play an essential role in unstable nuclei. Thus, it is natural to expect that the application of polarization techniques to RI-beam experiments would bring a better understanding of structure and dynamics of nuclei far from the stability line. \\
\begin{figure}[htbp]
 \centering
 \resizebox{8.5cm}{!}{\includegraphics{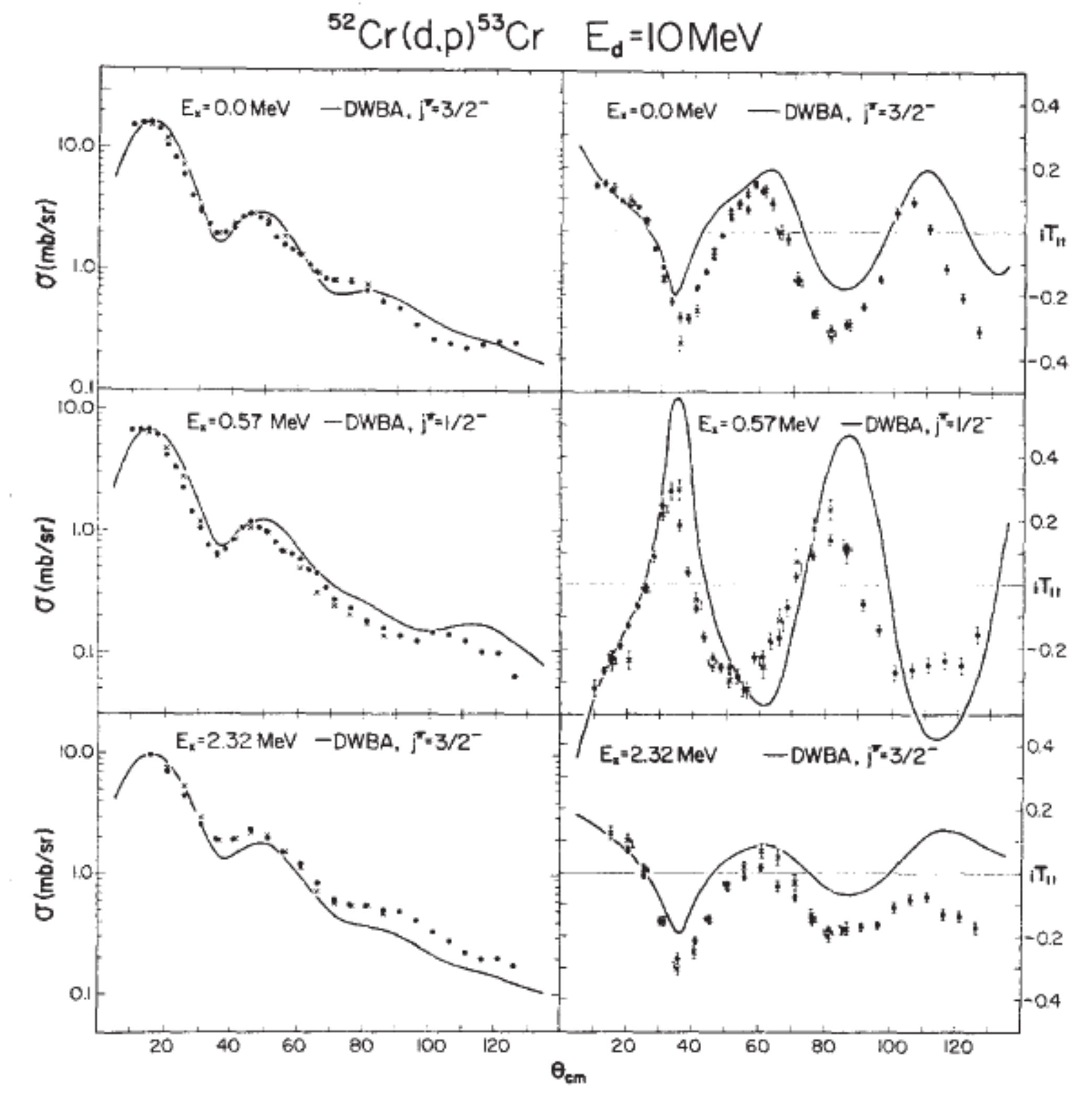}}
 \caption{(Left) Differential cross section to different final states in $^{53}$Cr from the one-neutron transfer $^{52}$Cr(d,p)$^{53}$Cr (open circles) at 10-MeV bombarding energy. (Right) Analysing power T$_{11}$ from the same reaction. Reprinted from~\cite{Kocher72}, with permission from Elsevier.\label{fig6}  }
\end{figure}
One of the most spectacular examples of the polarization effects can be found in vector analyzing power for the $(d,p)$ and $(p,d)$ reactions at low incident deuteron and proton energies of 10 MeV. Figure~\ref{fig6} shows the cross-section and vector analyzing power data for the ${\rm ^{52}Cr}(d,p){\rm ^{53}Cr}$ reaction at $E_{d}=10$~MeV~\cite{Kocher72}. The cross section for the states at excitation energies E$_x$=0, 0.57, 2.32~MeV are shown in left panels of Fig.~\ref{fig6}. Data exhibit angular distributions typical of $\Delta L=1$, where $\Delta L$ indicates transferred orbital angular momentum. The angular distribution for the transition to $E_{x}=0.57$~MeV levels with $j^{\pi}=1/2^{-}$ is almost identical to those for $E_{x}=0$ and 2.32~MeV levels with $j^{\pi}=3/2^{-}$. It is hardly possible with cross section data only to differentiate $3/2^{-}$ and $1/2^{-}$ levels.  On the other hand, a striking difference between $1/2^{-}$ and $3/2^{-}$ levels can be found in the vector analyzing power data. Namely, its sign is opposite depending on $j$, which makes the vector analyzing power a clear signature of $j$. This example clearly demonstrates how analyzing powers measured together with cross sections are essential for a complete spin-parity assignment. \\
Analyzing-power measurements have been proved to be useful also in the $(p,pN)$ knockout reactions~\cite{Kitching80} which are used at intermediate energy for the spectroscopy of hole states~\cite{Jacob66,Jacob73}. 
At incident energies higher than E=100~MeV, the reaction is usually dominated by a single-step nucleon-nucleon scattering and a single hole state is selectively populated without serious disturbance to the residual nucleus. Under these conditions, the distorted-wave impulse approximation (DWIA) can be reliably used to analyse these experiments.  \\
The use of polarized protons in knockout studies was first proposed by Jacob and Maris~\cite{Jacob76} and demonstrated at TRIUMF by Kitching {\it et al.}~\cite{Kitching80}. Figure~\ref{fig7} shows data of vector analyzing power (a) and cross section (b) for the ${\rm ^{16}O}(p({\rm pol}),2p)$ reaction at 200~MeV. Two protons in the final state were detected at 30$^{\circ}$ in the laboratory system. Filled and open circles represent data for knockout of protons in $p_{1/2}$ and $p_{3/2}$ states, respectively, as a function of the kinetic energy of one proton $E_{1}$. The $E_{1}$ dependence of the cross sections shown in Fig.~\ref{fig7}~(b)  is almost identical for $p_{1/2}$ and $p_{3/2}$ knockouts and it is obvious that cross section measurement alone does not allow us to make an unambiguous $J$ determination. In good contrast to this, the vector analyzing power data have opposite sign depending on $J$ and thus can be a clear signature of $J$.\\
\begin{figure}[htbp]
 \centering
 \resizebox{8cm}{!}{\includegraphics{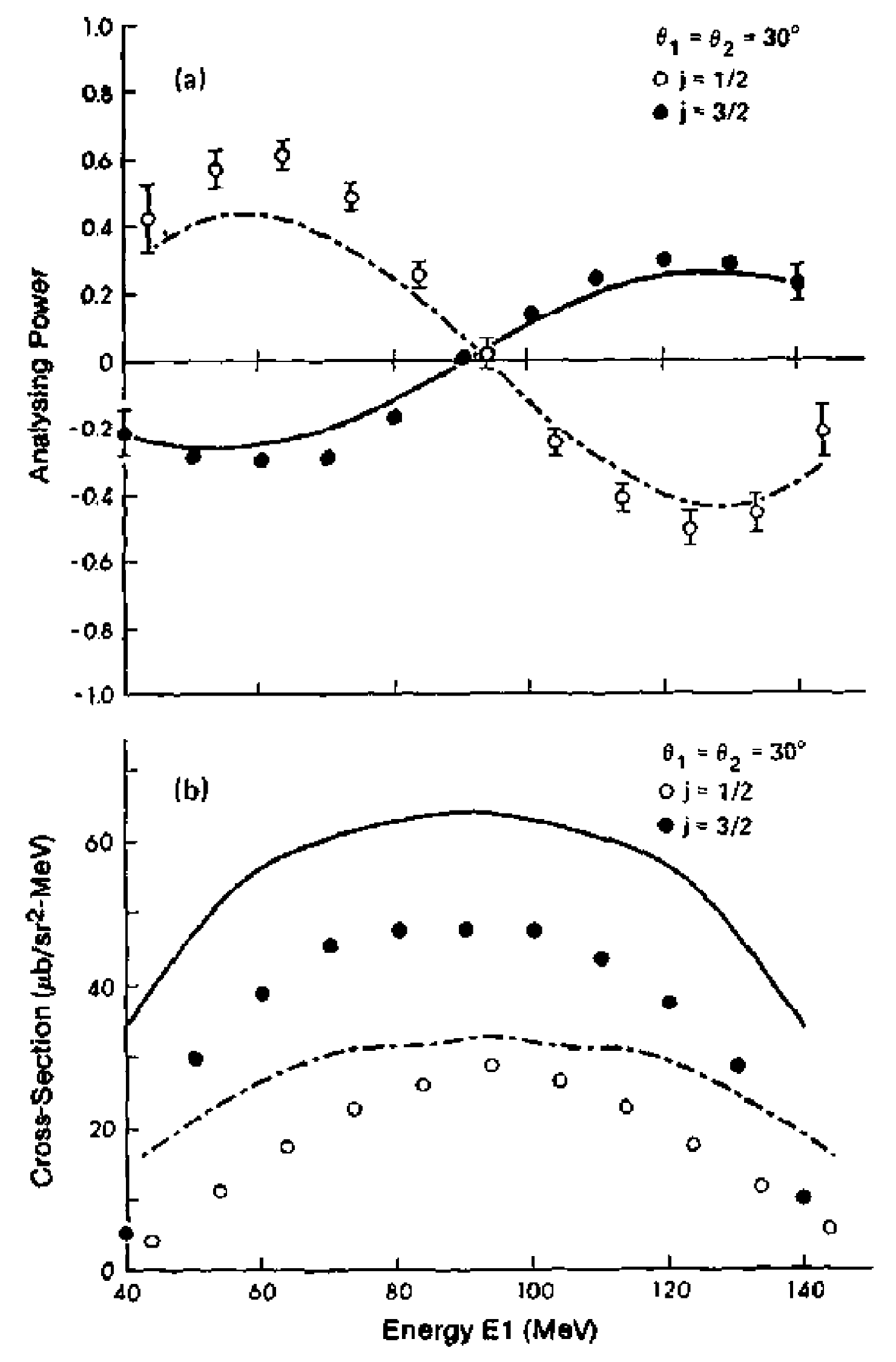}}
 \caption{The cross-section and vector analyzing power data in the ${\rm ^{16}O}(p,2p)$ reaction at $E_{p}$=200~MeV. Reprinted from~\cite{Kitching80}, with permission from Elsevier.} 
\label{fig7}
 \end{figure}
Polarization in proton elastic-scattering experiments is another interesting example. The measurement of vector analyzing power provides a unique way to determine the spin-orbit part of the optical potential. The spin-orbit potential is considered to be proportional to the derivative of the density distribution.  In nuclei with a large neutron excess, peculiar neutron distributions such as halo and skin exist. It is natural to expect the spin-orbit potentials in those neutron-rich nuclei are different from those of stable nuclei. Analyzing power measurement for the proton elastic scattering of neutron-rich nuclei is of interest in this respect. It should be noted that the proton elastic scattering at lower energies of several MeV/nucleon can be a good spectroscopic tool to single-particle resonant states. Here, an analyzing-power measurement is useful again to perform an unambiguous spin-parity assignment.\\

In the following, we present an overview of the last 10 year developments in cryogenic hydrogen targets dedicated to the study of exotic nuclei in inverse kinematics. We first focus in section~\ref{section1} on non-polarized targets, either liquid or solid. An overview of recent efforts in  gas targets for transfer and inelastic scattering is also given to a lesser extent. The developments in polarized targets recently achieved are presented in  section~\ref{section2}. A summary of experiments performed with such setups is highlighted to illustrate the physics approached with the presented devices. All targets discussed in this review and their domain of application are summarized in Table~\ref{tab1}.

\begin{table*}
\begin{center}
\begin{tabular}{|l|l|c|c|c|c|c|}
\hline
Phase&Target&(p,p),(p,p')$^{\dagger}$&(p,X$\gamma$)&(p,pN)$^{\dagger}$&N-transfer$^{\dagger}$&(p,n)\\
\hline
Gas&TPCs (ex. MAYA)&$\bigcirc$&$\times$&$\times$&$\bigcirc$&$\triangle$\\
\hline
Liquid&GANIL liquid&$\times$&$\triangle$&$\times$&$\times$&$\triangle$\\
\hline
&CRYPTA&$\times$&$\bigcirc$&$\times$&$\times$&$\bigcirc$\\
\hline
&PRESPEC&$\times$&$\bigcirc$&$\times$&$\times$&$\triangle$\\
\hline
&MINOS$^{\star}$&$\times$&$\bigcirc$&$\bigcirc$&$\times$&$\times$\\
\hline
Solid&RIKEN solid&$\times$&$\bigcirc$&$\bigcirc$&$\times$&$\triangle$\\
\hline
&GANIL solid&$\bigcirc$&$\bigcirc$&$\times$&$\triangle$&$\triangle$\\
\hline
&RIKEN ultra-thin$^{\star}$&$\bigcirc$&$\triangle$&$\bigcirc$&$\triangle$&$\triangle$\\
\hline
&CHyMENE$^{\star}$&$\bigcirc$&$\triangle$&$\triangle$&$\bigcirc$&$\triangle$\\
\hline
&CNS (pol)&$\bigcirc$&$\triangle$&$\bigcirc$&$\times$&$\bigcirc$\\
\hline
&ORNL-PSI (pol)&$\bigcirc$&$\triangle$&$\triangle$&$\bigcirc$&$\triangle$\\
\hline
\end{tabular}
\caption{Domains of use for the different targets presented in this review. $\bigcirc$: designed for this type of reactions, $\triangle$: can be used, $\times$: difficult or impossible to apply. Reactions marked with a $\dagger$ are understood as missing mass measurements. The GANIL liquid target was dedicated to reaction cross section measurements. Targets marked with a $\star$ are projects in development.}
\label{tab1}
\end{center}
\end{table*}

\section{Non-polarized targets}
\label{section1}
\subsection{Solid targets}
\subsubsection{The GANIL target}
\begin{figure}[t]
\begin{center}
\includegraphics[trim= 40mm 20mm 50mm 10mm,clip,width=7cm]{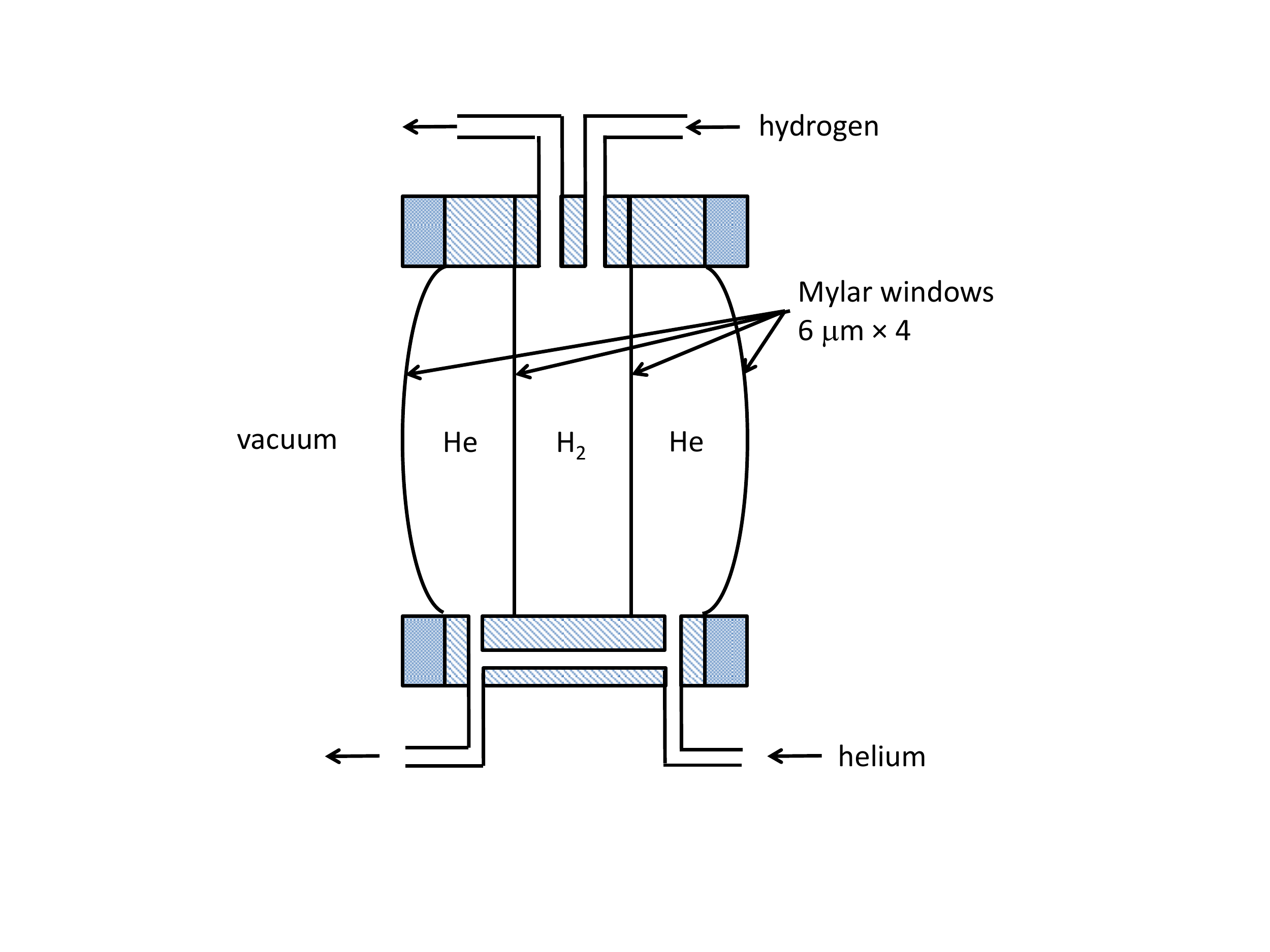}
\caption{Scheme of the GANIL solid hydrogen target. Homogeneous thicknesses down to 1 mm have been achieved. Double windows are used to keep the target thickness constant during the cooling process.}
\label{fig8}
\end{center}
\end{figure}
The GANIL target~\cite{Dolegieviez200632} was developed to perform transfer or resonant elastic scattering experiments at incident energies lower than 10 MeV/nucleon. A thin target with a rather high density of scattering centers was foreseen and therefore the development of a new solid target with a thickness ranging from 1 to 5 mm and a high homogeneity was undertaken. For high-density hydrogen targets, either liquid or solid targets can be used. The filling of the target implies high pressures and, in the case of H$_2$, at least 100 mbar are necessary to remain above the triple point. To avoid windows deformations that would result in inhomogeneous targets, a transition to the liquid phase (16.2 K/230 mbar) before a progressive solidification (T$\leq$13.9 K) was chosen. Liquid helium is used as a cold source at 4 K and the growth of the target is imposed by the temperature gradient in the metal frame supporting the target.\\

\begin{figure*}
\begin{center}
\includegraphics[trim=0mm 10mm 20mm 10mm,clip,angle=-90,width=8.cm]{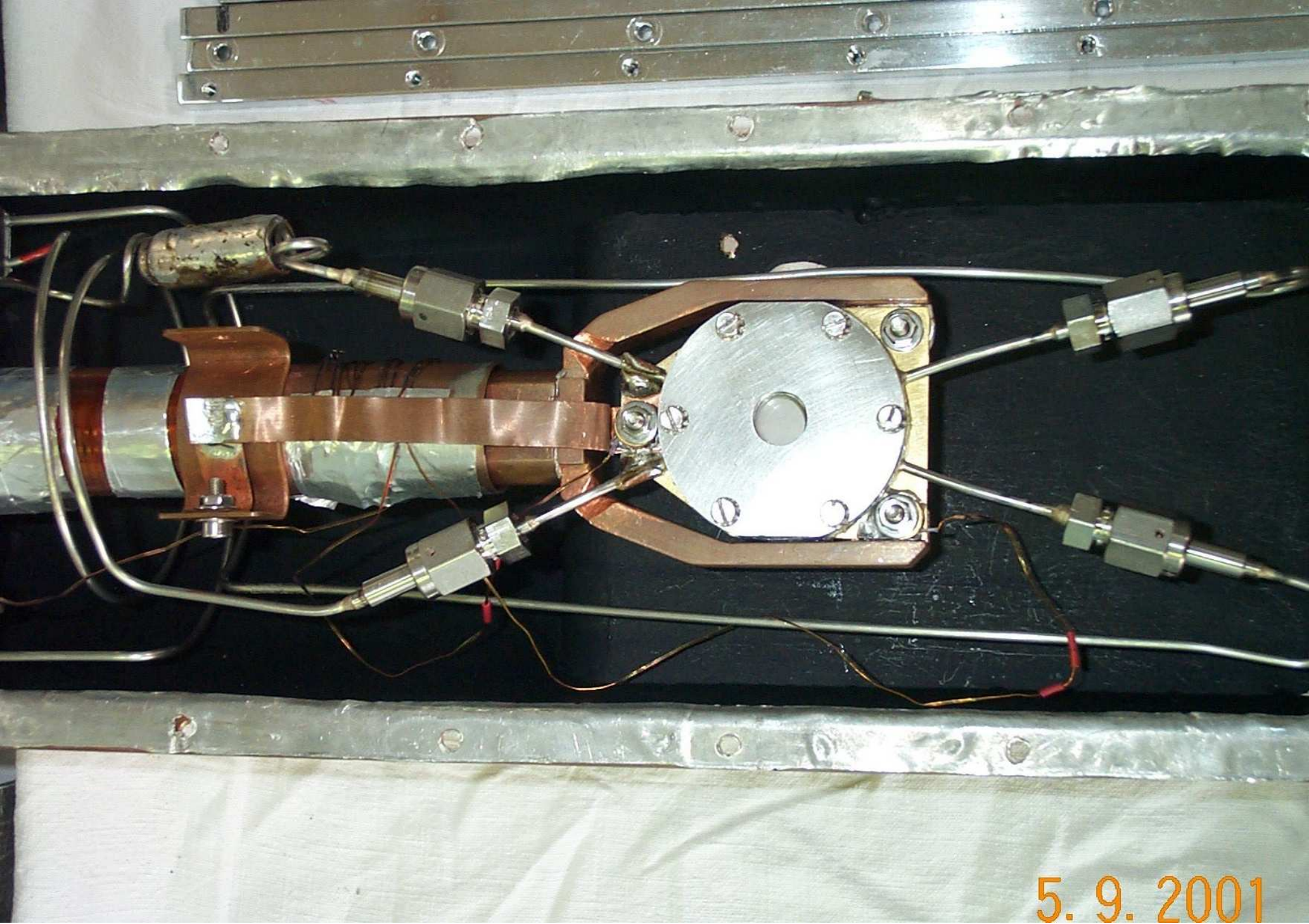}
\includegraphics[trim= 20mm 10mm 0mm 10mm,clip,angle=-90,width=8.cm]{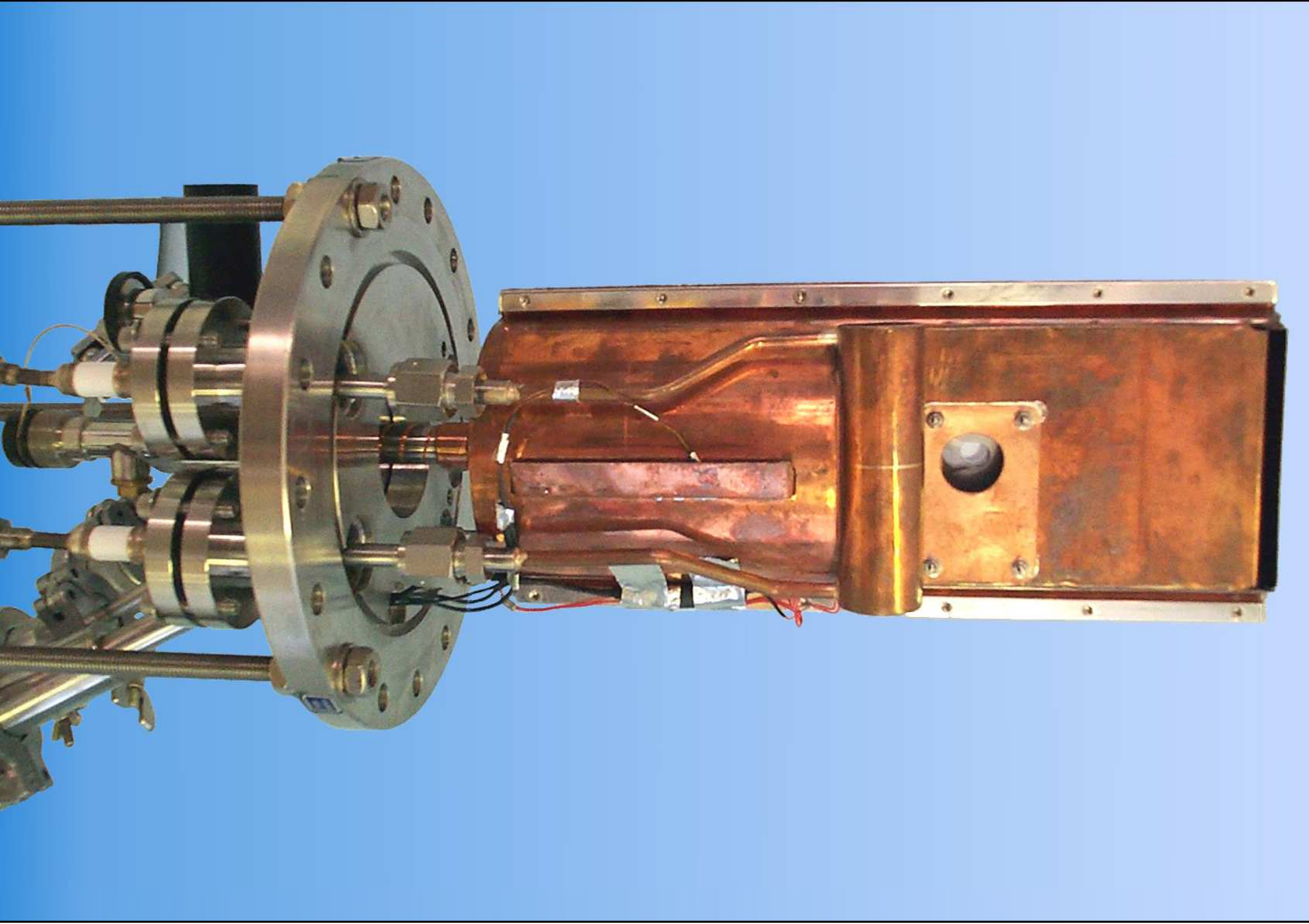}
\caption{(Color online) Pictures of the GANIL solid hydrogen target. (Left) Open target ensemble. The Helium tubes for the side pockets are located on the bottom of the target, the hydrogen fluid enters the target through the capillar tubes at the top. The diameter of the target is 10 mm. (Right) Complete target with its copper thermal screen. The cylinder copper volume on top of the target cell is filled with liquid N$_2$ and acts as a cold source. Courtesy P. Dolegieviez (GANIL).}
\label{fig9_10}
\end{center}
\end{figure*}

\begin{figure}
\begin{center}
\includegraphics[trim= 50mm 10mm 50mm 10mm,clip,angle=-90,width=6cm]{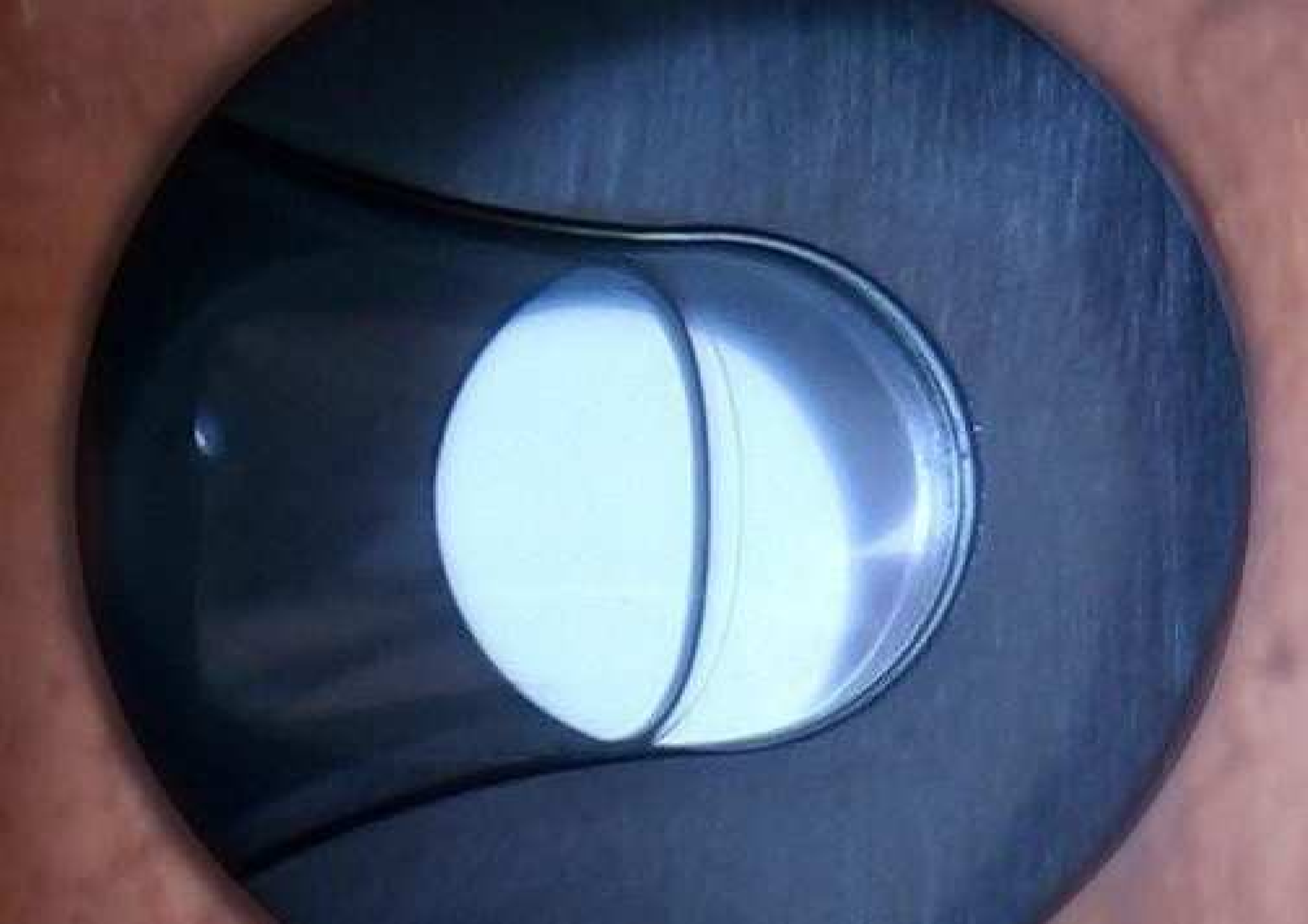}
\caption{(Color online) Picture of the GANIL solid hydrogen target cell during the cooling process. The three phases of hydrogen are visible from bottom to top: solid, liquid and gas. Courtesy P. Dolegieviez (GANIL).}
\label{fig11}
\end{center}
\end{figure}

The principle of the target is based on the use of double Mylar windows (see scheme in Fig.~\ref{fig8}). The target is made of a metal frame to which Mylar windows are glued. A stack of frames forms an H$_2$ or D$_2$ target cell with a He cell on both sides of the target. During the target production phase, identical pressure is applied on both sides of the target windows, in order to maintain the inner windows free from pressure constraints. To do so, the pressure of the Helium gas volume matches the pressure variations of the hydrogen circuit during the phase transitions and up to the complete solidification of the target body. Once the solid is formed, helium gas is taken out. The mechanical strength of the windows of the helium circuitry with respect to the beam vacuum imposes the filling pressure, and hence the phase change temperature of the target gas. The total windows thickness used is four times 6 $\mu$m of Mylar (3.3 mg/cm$^2$). The thickness of the target is determined by the mechanics of the support of the inner windows. A minimum of 0.5 mm seems achievable, with a very homogeneous thickness along the surface of the target. Such a thickness would require the smaller target diameter. Up to now, the minimal thickness that has been used is 1 mm, which corresponds to 7.5 mg/cm$^2$.

The preparation of a target lasts about 4 hours:
\begin{itemize}
\item The target is mounted in the reaction chamber. The vacuum is set simultaneously in all cells of the target (hydrogen and helium) and in the vacuum chamber not to destroy the windows. 
 \item The target cell and the external pockets are filled with hydrogen and He, respectively. The filling is made at room temperature from 10$^{-1}$ mbar to 230 mbar. The filling takes approximately 10 minutes.
\item Liquid nitrogen is transferred to the reservoir in contact with the 80 K copper shield surrounding the target (see Fig.~\ref{fig9_10}). The shield is set to a temperature of 77 K. This operation lasts 1 hour.
\item Liquid helium at 4 K is transferred via a transfer pipe to the cold finger thermally coupled to the bottom of the target frame. The bottom of the target frame reaches 5 K nearly 35 minutes after the He transfer, while the top of the target is still at 27 K. At that point, the bottom part of the target cell is at 16.2 K and hydrogen liquid starts forming (see Fig.~\ref{fig11}). The target is entirely solid 2 hours after He is introduced. All along the cooling process, the He gas pressure is controlled and maintained at the same pressure than the hydrogen, following its variations due to phase changes. Hence, the windows of the target do not deform: He pockets from both sides of the target act like a mold.
\end{itemize}

Two experiments have been performed at GANIL by using this device:
\begin{itemize}
\item  $^{18}$Ne(p,p)$^{18}$Ne at 7.2 MeV/u, GANIL~\cite{Oliveira}. The excitation function of $^{19}$Na from the elastic-scattering reaction  was measured with the first radioactive beam from the SPIRAL facility at the GANIL laboratory. Several broad resonances have been observed, corresponding to new excited states in the unbound nucleus $^{19}$Na.
\item $^{26}Ne$(d,p$\gamma$)$^{27}$Ne at 10 MeV/nucleon, GANIL~\cite{OBE063,PhysRevC.74.064305}. Inclusive measurement coupled to EXOGAM for $\gamma$-ray detection and the magnetic spectrometer VAMOS for residue identification. Two transitions have been observed in $^{27}$Ne. A low-lying negative parity state (intruder from the $fp$ shell) has been suggested from the data showing that $^{27}$Ne is at the edge of the so-called N~=~20 island of inversion.
\end{itemize}

\begin{figure*}[htbp]
\begin{center}
\includegraphics[width=12cm]{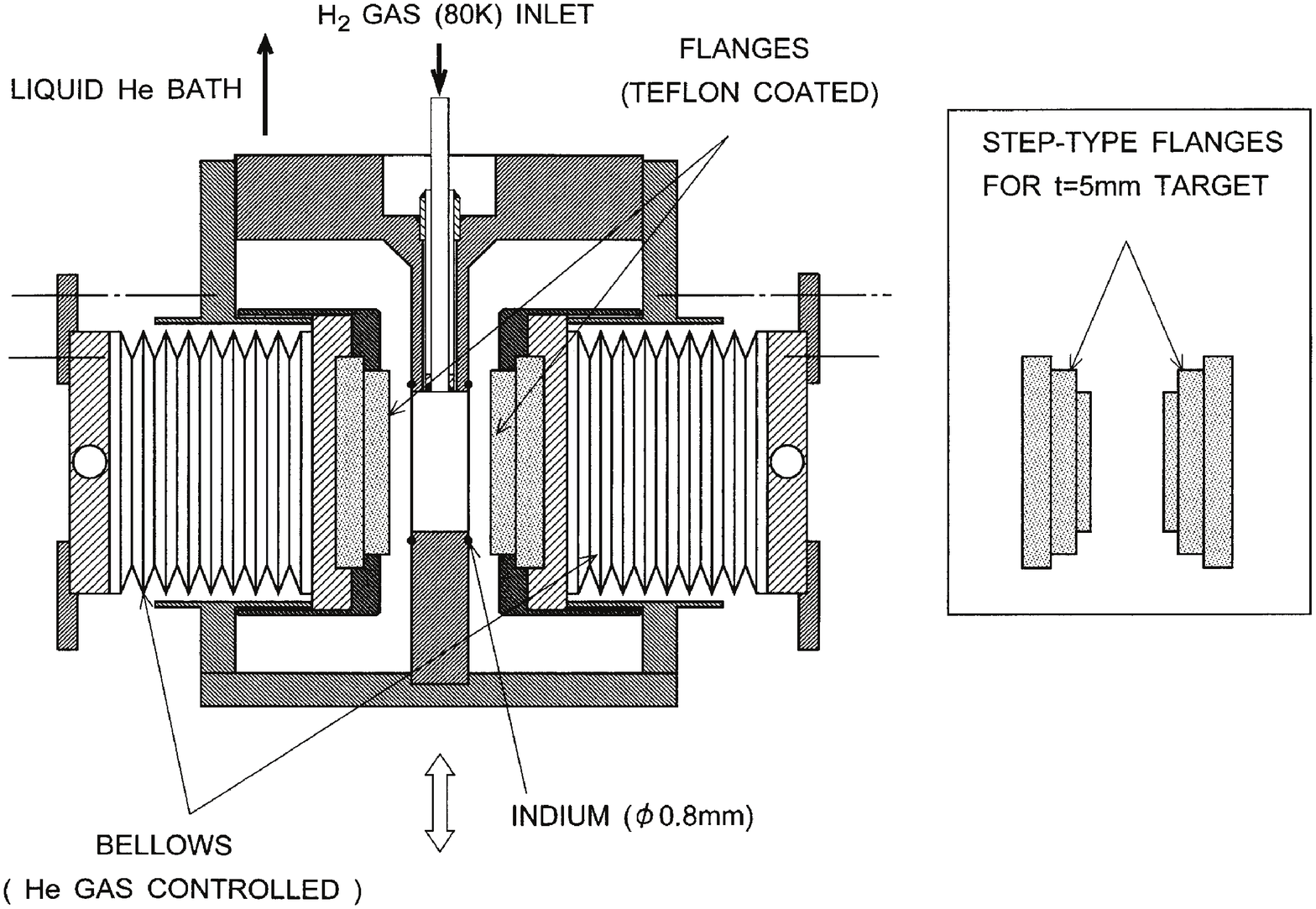}
\end{center}
\caption{Schematic view around the windowless target.
Copper cell system and mechanical press consisting of
polytetrafluoroethylene-coated stainless steel flanges 
and stainless steel bellows (left).
Polytetrafluoroethylene-coated flanges to fabricate 5-mm-thick targets (right). Reprinted from~\cite{Ishimoto2002304}, with permission from Elsevier.
}
\label{fig12}
\end{figure*}

\subsubsection{Solid Hydrogen Target for Recoil detection In Coincidence with Inverse Kinematics (SH TRICK)}
\label{RIKENsolidTarget}
Solid targets suitable for missing mass spectroscopy
at intermediate energies (200--300 MeV/$u$) have been developed
by a collaboration of Japanese universities and accelerator facilities.
The original target was studied by a collaboration of Tohoku University and KEK 
so that single-particle properties of exotic nuclei could be obtained
from experiments of quasi-free ($p$,$pN$) reactions in inverse kinematics~\cite{Ishimoto2002304}.\\

The outline of the first developed windowless solid target is as follows.
H$_{2}$ crystals were grown reproducibly
in a cell bored in a 10-mm-thick pure copper plate
which was in direct contact with a liquid He reservoir.
In order to obtain uniform surfaces,
two polytetrafluoroethylene-coated stainless steel flanges were pressed tightly
on the surfaces of the plate by stainless steel bellows 
which were pressurized by helium gas (see Fig. \ref{fig12}).
H$_{2}$ gas was supplied through a stainless-steel tube to the cell
where it was crystallized between temperatures of 4.7 and 7.3 K.
The diameter of the crystal was 25 mm.
The thickness was chosen to be either 5 or 10 mm 
from two kinds of polytetrafluoroethylene-coated flanges.
After crystallization,
the mechanical press was separated from the plate
and removed to a remote position inside a cryostat. 
The crystal was self-supported in the cell without any extra material in its neighborhood. 
The sublimation loss is negligibly small if the crystal is held at 3K.

Based on this first development, another cryogenic system has been built.
A collaboration of RIKEN, Kyoto University, and the above-mentioned group
also constructed the same one to measure elastic scattering of protons with radioactive ion beams. The project is called ESPRI.\\
The difference from the previous system is as follows.
In order to use the target in several laboratories and perform long-term experiments,
the device uses a two-stage Gifford-McMahon refrigerator
(Sumitomo Heavy Industries, RDK-415) rather than liquid He.
The second stage of the refrigerator
has a refrigeration capacity of 1.5 W at a temperature of 4.2 K.
The size of the target was 30--40 mm in diameter and 5 mm in thickness 
in the experiment of the ($p$,$pN$) reaction
while it was 30 mm in diameter and 1 mm in thickness in the other experiment.
The size is desirable because the beam generally has a large size and a low intensity.
The target was surrounded by a copper radiation shield.
The shield was attached to the first stage of the refrigerator
that had a temperature of below 50 K.
The shield had not only two 30-mm-diameter beam windows for beams
but also a large opening for protons.
Since the openings to areas at room temperature
cause serious thermal radiation,
sublimation occurs around the center of the target.
Therefore, 9-$\mu$m-thick aramid films were used to cover the cell;
they were attached to the pure copper plate by epoxy resin (Stycast,1266).
The backgrounds from the film material are negligible compared with
the protons from solid H$_{2}$.

\begin{figure}[t]
\begin{center}
\includegraphics[width=7cm]{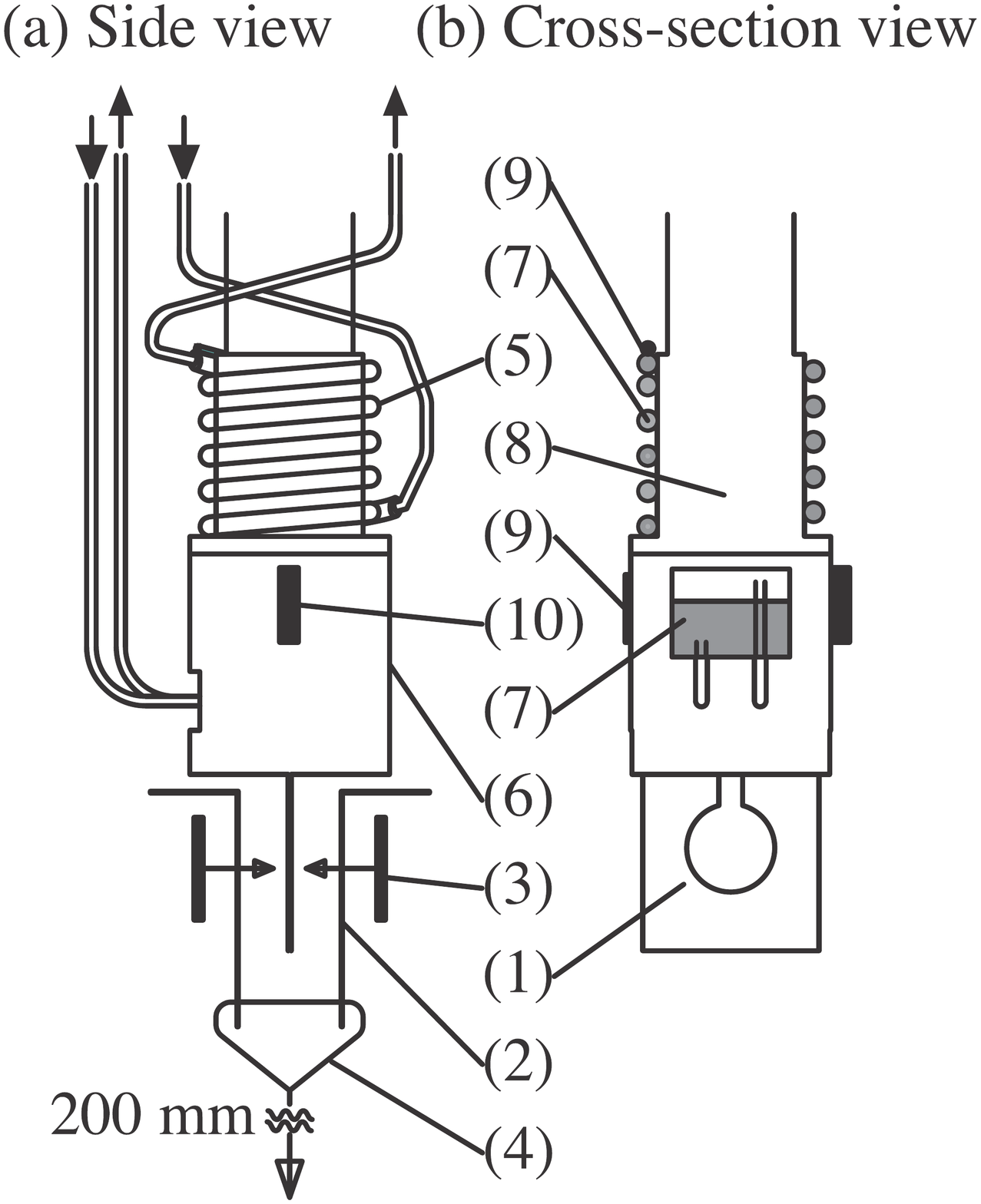}
\end{center}
\caption{Schematic view around the target using para-H$_{2}$.
(a) Side and (b) cross-section views are shown.
(1) Target cell;
(2) aluminum plate; 
(3) neodymium magnet;
(4) cotton thread;
(5) ortho--para (pipe) converter;
(6) ortho--para (cylinder) converter;
(7) catalysis (iron(I\hspace{-.1em}I\hspace{-.1em}I) oxide-hydroxide);
(8) second stage of refrigerator;
(9) Pt--Co thermometer;
(10) heater.}
\label{fig12b}
\end{figure}

Unfortunately,
it has been difficult to fabricate a target 
that is 30 mm in diameter and 1 mm in thickness
due to the thermal problems in the system.
A new development has been recently undertaken~\cite{matsuda2011}.
The main limitation to fabricate the target
was a too low thermal conductivity of normal H$_{2}$
to stand the radiative warming from the environment.
The thermal conductivity of solid H$_{2}$ increases
using highly concentrated para-H$_{2}$ that 
has a thermal conductivity of 100 W/m$\cdot$K at a temperature of 4 K,
which is more than 100 times larger than that of normal H$_{2}$. 
At room temperature, normal H$_{2}$ gas is composed of 25\% of para-H$_{2}$ and
75\% of ortho-H$_{2}$. 
The ratio of para-H$_{2}$ increases at lower temperature 
but the conversion to an equilibrium state is very slow without a catalyst.
The new development used
Iron (III) oxide (Aldrich, hydrated, catalyst grade, powder, 30--50 mesh)
as a catalyst to enhance the conversion of para-H$_{2}$ (see Fig. \ref{fig12b}).
The obtained concentration of para-H$_{2}$ was close to 100\%.
Finally, a homogeneous pure H$_{2}$ target has been successfully fabricated
using para-H$_{2}$ and a simple mechanical press
consisting of aluminum plates and neodymium magnets.
An in-beam measurement with $^{10,11,12}$C beams at 300 MeV/nucleon
has been performed during 180 hours with no observable
characteristics change in the hit region of the target.
This system was also installed in the other target recently. The solid target for ESPRI is named as SH TRIC\begin{sideways}\begin{sideways}K\end{sideways}\end{sideways} (Solid Hydrogen Target for Recoil detection In Coincidence with Inverse Kinematics). The system using para-hydrogen is called SpH TRIC\begin{sideways}\begin{sideways}K\end{sideways}\end{sideways}.

\subsubsection{The CHyMENE project}
The CHyMENE project (french acronym for "Cible Mince d'Hydrogene pour l'Etude des Noyaux
Exotiques", \emph{i.e.} "Thin Hydrogen Target for the Study of Exotic Nuclei") aims at providing
a very thin windowless hydrogen target down to thicknesses lower than  50 microns~\cite{gillibert}. The technique does not present any upper limit for the target thickness, except for the cryogenic power avalaible, and millimetric targets could also be conceived. The project is at an R\&D phase and is carried out at CEA Saclay. The production of the target is based on the
extrusion technique commonly used in plasma physics to produce ice-hydrogen pellets to fuel Tokamaks~\cite{PhysRevLett.42.97,PhysRevLett.53.352}. The following presents the technique and first results obtained with a prototype.\\
\begin{figure}[!]
\begin{center}
\includegraphics[trim= 35mm 100mm 35mm 70mm,clip,width=8 cm]{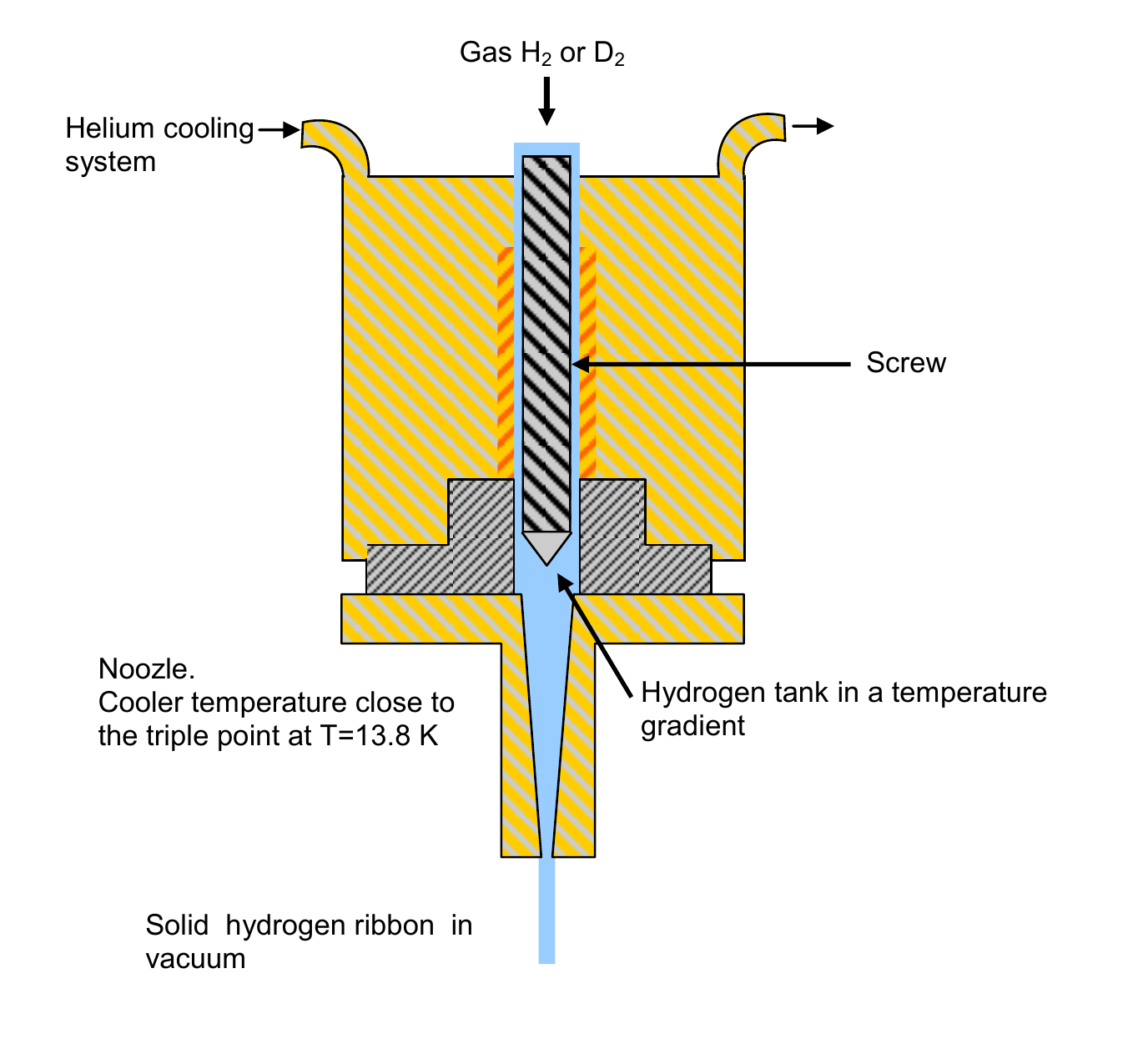}
\caption{Scheme of the extrusion technique of the CHyMENE project.}
\label{fig13}
\end{center}
\end{figure}

In this technique, a screw is used to push hydrogen towards an extrusion nozzle in a continuous way. Hydrogen gas inside the extruder is cooled down with He gas near the triple point (T$_T$ = 13.8 K, P$_T$ = 70.4 mbar, see Fig.~\ref{fig1}) in an amorphous phase. At the nozzle, hydrogen is extruded through a rectangular cross section noozle which is used to define the target thickness (see Fig.~\ref{fig13}). The hydrogen film is pushed by the screw in the reaction chamber at a velocity of about 1 cm.s$^{-1}$. The  H$_2$ sheet is partially protected from background radiation by a thermal screen.
So far, solid hydrogen films have been obtained with a good reproducibility for two thicknesses, 200 and 100 $\mu$m. A positive result has been recently obtained for a 30 $\mu$m thickness. Laboratory target extrusions were made to study production conditions and a first in-beam test with a low-energy proton beam and elastic scattering was done.  \\
\begin{figure}[!]
\begin{center}
\includegraphics[trim= 0mm 0mm 0mm 0mm,clip,width=7.5cm]{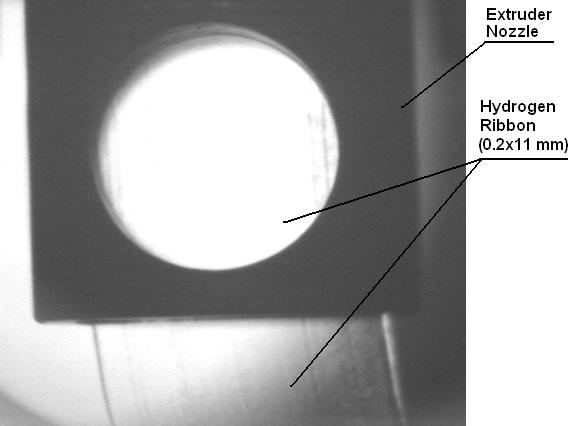}
\caption{Picture of a 200-$\mu$m thick H$_2$ film extruded at 10 mm.s$^{-1}$ in a reaction chamber. The film width is 11 mm.}
\label{fig14}
\end{center}
\end{figure}

A test to study the target production was done with a nozzle whose thickness was 100 $\mu$m. The device is cooled down by He gas from room temperature to 10 K. Then, hydrogen gas is introduced in the extruder volume. When the volume of the extruder is full of H$_2$ ice, the rotation of the screw can start.  A good vacuum level better than 10$^{-4}$ mbar is needed in the chamber for stable extrusion. During extrusion, the pumping is also necessary to remove H gas due to sublimation  from the reaction chamber. The temperature at the nozzle is stabilized around 17 K. The hydrogen film is optically transparent (see Fig.~\ref{fig14}).\\

Both thickness and homogeneity have been investigated in a dedicated in-beam characterization.The target properties have been investigated via proton-proton elastic scattering where scattered protons were detected at a given angle $\theta$. The number of scattered protons $N_{scatt}(\theta,t)$ in the detectors is proportional to the thickness of the target and its variations with time $t$. The variance of the proton energy distribution is sensitive to the homogeneity of the target. 

The test experiment was done with a proton beam at E$_p$ = 3 MeV from the tandem accelerator at CEA/DAM/DIF. The extruder device was installed on a beam line, so that the hydrogen target could be irradiated continuously by the beam. Silicium detectors with collimators were located downstream at 12 degrees in the laboratory frame inside the reaction chamber.

\begin{figure*}[t]
\begin{center}
\includegraphics[trim= 70mm 0mm 0mm 0mm,clip,angle=90,width=15cm]{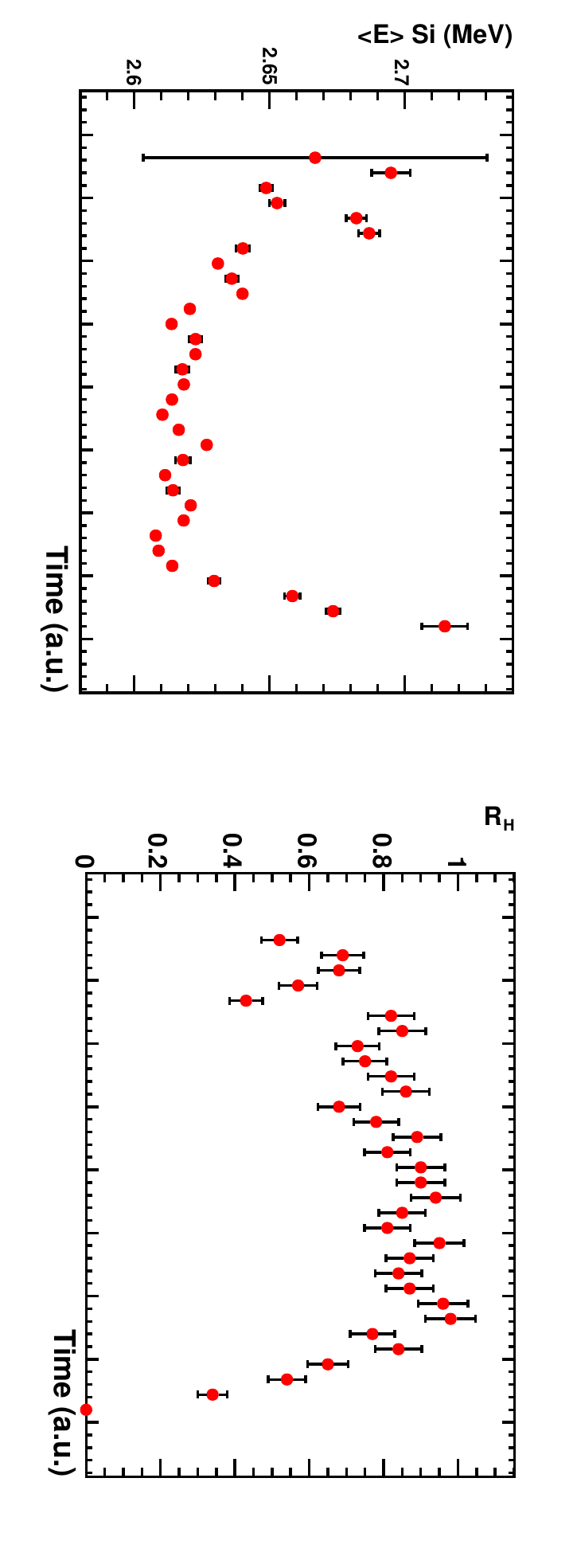}
\caption{(Color online) (Left) Energy spectrum of scattered protons measured during the in-beam test of the CHyMENE prototype as a function of time. The measured energy is directly linked to the target thickness. The time range of the displayed measurement is 2 minutes. (Right) Ratio R$_H$ (see text) as a function of time.}
\label{fig14bis}
\end{center}
\end{figure*}

A hydrogen target was produced successfully with thickness 100 $\mu$m in that environment. However, the vacuum level without extrusion was not good enough to get the same conditions during extrusion as previously. During stable conditions, the thickness was stable corresponding to a scattered proton with energy between 2.55 and 2.6 MeV (see left part of Fig.~\ref{fig14bis}), in agreement with kinematics and energy loss calculations that predict 2.66 MeV at the entrance of a detector placed at 12 degrees from the beam axis with a target thickness equal to 100 $\mu$m. The measured number of scattered protons by unit of time (N$_{exp}$) has been compared to the expected number of counts assuming a 100 $\mu$m target (N$_{th}$). The ratio $R_H=N_{exp}/N_{th}$ is plotted on the left panel of Fig.~\ref{fig14bis} and shows a value consistent with 1 during stable conditions.\\
The energy width of the protons measured during the measurement has been compared to simulations taking into account the instrinsic energy resolution of the setup, energy spread inside the target, energy straggling and angular aperture of the collimators. The measured width can only be reproduced if one assumes inhomogeneities in the thickness of about 12 $\mu$m or 12\% ($\sigma$ value) around 100 $\mu$m. This first test gives confidence that the production of windowless targets of 100 $\mu$m, and maybe less, could be obtained with this technique with a thickness homogeneity better than 10\%. Further work on the extrusion conditions and surface treatment of the nozzle are requested to achieve a proper level of thickness homogeneity.

\subsubsection{RIKEN ultra-thin solid target}
With the advent of high-intensity RI-beams and sophisticated particle detectors, missing mass spectroscopy in the inverse kinematics is becoming a powerful tool in RI-beam experiments. Here a new demand arises for a hydrogen target dedicated to $(p,p')$ reaction studies in inverse-kinematics. In these experiments, protons scattered at forward angles in the center-of-mass system have very low energies. To carry out the measurement of protons scattered at small center-of-mass angles, a thin and pure hydrogen target is suitable. \\

A new solid hydrogen target is being planned at RIKEN\cite{Mizuki11}: the goal is to realize a target with a uniform thickness of 1~mg/cm$^{2}$, for detecting 0--10~MeV protons with a sufficient angular-resolution. The target has an advantage over a propylene target due to the significant increase in the number of hydrogen atoms for the same thickness. It should be also noted that, today, there is no windowless system producing such low thicknesses of solid hydrogen. 

The most specific feature of the target is the use of nano-membranes as target windows. The target cell has three layers: the central layer with a geometrical thickness of 0.12~mm contains hydrogen. In the cooling-down process, the layer is sandwiched by outer layers filled with helium gas at the same pressure of hydrogen to avoid swelling of the hydrogen layer. After hydrogen is solidified, the helium-gas, together with windows dividing the gas from vacuum is removed. Finally one get a 1~mg/cm$^{2}$ solid hydrogen target sealed by the nano-membranes. The nano-membrane developed at RIKEN\cite{Watanabe07} has a thickness of  25--200~nm. 

\subsection{Thick liquid hydrogen targets}
\subsubsection{The CRYPTA liquid hydrogen target}
The CRYPTA target is based on an original development at Kyushu university by K. Sagara and H. Ryuto for experiments at RCNP laboratory. The present target has been developed for nuclear-physics experiments at RIKEN~\cite{Ryuto20051}. The typical thickness is 30 mm. The target has been extensively used to perform the spectroscopy of exotic nuclei at the RIPS separator.\\
An Aluminium target cell is attached to the second stage of a Gifford-MacMahon cycle refrigerator through a copper connecting component. The cooling capacities of the first and second stages of the refrigerator are 30 W at 77 K and 10 W at 20 K, respectively. Havar foils are glued with epoxy resin onto the aluminium rings attached to the cylindrical target cell with indium gaskets forming a 24-mm diameter entrance and exit windows. In some experiments, a target cell that has a 30-mm diameter can be used.
\begin{figure*}[t]
\begin{center}
\includegraphics[trim= 60mm 55mm 60mm 60mm,clip,width=15cm]{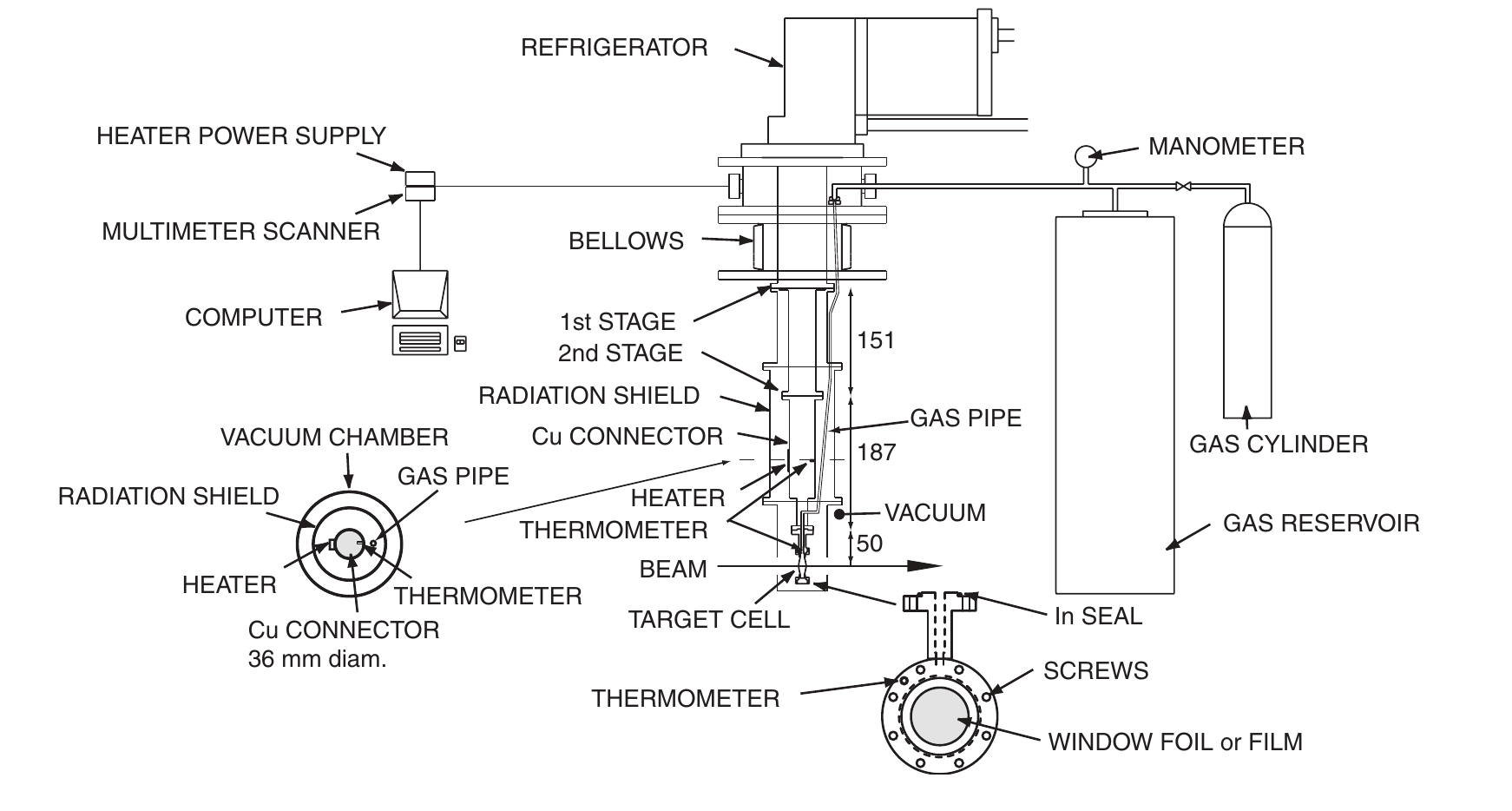}
\caption{General scheme of the CRYPTA liquid-hydrogen target from RIKEN. Reprinted from~\cite{Ryuto20051}, with permission from Elsevier.}
\label{fig15}
\end{center}
\end{figure*}
Typical thicknesses of 6 $\mu$m for the Havar foils are used. The windows exhibit sufficient stability and vacuum tightness. The maximum internal pressure at room temperature is approximately 0.4 MPa when 12.5 $\mu$m-thick aramid films are used for 24-mm diameter windows. The 6-$\mu$m-thick havar foil that forms the 40-mm diameter window expands of about 1 mm at the top when the foils experience a 0.1 MPa pressure difference. The target cell is connected to a 10-liter gas reservoir. When a thicker target is requested a larger reservoir of 30 liters is used. The hydrogen gas is cooled down to aproximately 18 K while passing through the connecting component, liquefied, and accumulated in the target cell. The gas is sealed in a closed system in order to use expensive deuterium gas effectively and for safety reasons. Consequently, the gas pressure measured with a manometer decreases steeply when the gas is liquefied. The target cell and the reservoir are always connected in order to prevent overpressure in the target cell caused by any accidental vaporization. It is also important for safety reasons to prevent air leakage into the gas system, and to use pipes with a sufficiently large internal diameter so as not to be clogged by a small quantity of air mixed into the hydrogen gas.  The temperature of the cryogenic section is controlled with a heater attached to the connecting component. A typical 24-h fluctuation of the gas pressure is 1\%. The typical working temperature are 18 K and 21 K, which are approximately 1.5 K lower than the boiling points of hydrogen and deuterium at 80 kPa, respectively.  The target is always free of bubbles and of constant thickness. The target cell, the connecting component, and the second stage of the refrigerator are surrounded by a 3-mm thick aluminium radiation shield connected to the  first stage of the refrigerator. The shield has two 50-mm diameter holes for the beam entrance and exit.\\
It takes about 100 minutes to fill the target with liquid hydrogen after the refrigerator begins the cooling from room temperature. \\
\begin{figure}
\begin{center}
\includegraphics[trim= 0mm 0mm 0mm 0mm,clip,width=8cm]{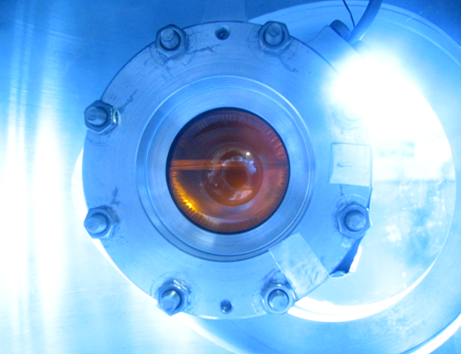}
\caption{Front photograph of the CRYPTA target.}
\label{fig15bis}
\end{center}
\end{figure}
Numerous gamma spectroscopy experiments have been performed by coupling this device to a gamma-ray detector such as DALI2 at RIKEN or NaI detectors at the NSCL. Proton or deuteron inelastic scattering measurements with these setups have provided important information on collectivity in neutron-rich nuclei:
 \begin{itemize}
\item The collectivity and decoupling of neutron and proton excitations has been studied along the carbon isotopic chain~\cite{PhysRevC.73.024610,Satou2008320,PhysRevC.79.011302}.  This phenomenon is observed for carbon isotopes heavier than $^{14}$C (see Fig.~\ref{fig4}).
\item The region of the N=20 island of inversion has been studied for the most exotic Ne, Na, Mg and Si isotopes~\cite{PhysRevLett.96.182501,Yanagisawa200384,PhysRevC.73.044314,NSR2009TA08,NSR2003IW02} isotopes produced at RI facilities.
\item Shell gap evolution at N=28 in $^{36,38,40}$S(p,p')~\cite{Campbell2007169,Campbell2007272}, and at Z=28 and N close to 50 in $^{74}$Ni(p,p')~\cite{Aoi2010302}, and collectivity at N=40 in neutron rich Cr isotopes~\cite{PhysRevLett.102.012502} were investigated via proton inelastic scattering.
\end{itemize}

CRYPTA is also used to measure reaction cross sections~\cite{NSR2006RO34} and to perform $(p,n)$ charge exchange~\cite{NSR2011SA02} and nucleon removal~\cite{PhysRevC.79.014602,NSR2010EL05,NSR2010NI10} studies.\\

A new development with a similar design inspired from the RIKEN target is being developed at Ursinus university (USA). The target was successfully commissioned in a $(p,n)$ charge-exchange experiment at the NSCL in october 2010. Two target-cell thicknesses  are now available for experiments: 7 and 30 mm.\\

\subsubsection{The GANIL liquid target}
\begin{figure}
\begin{center}
\includegraphics[trim= 0mm 30mm 0mm 0mm,clip,width=7cm]{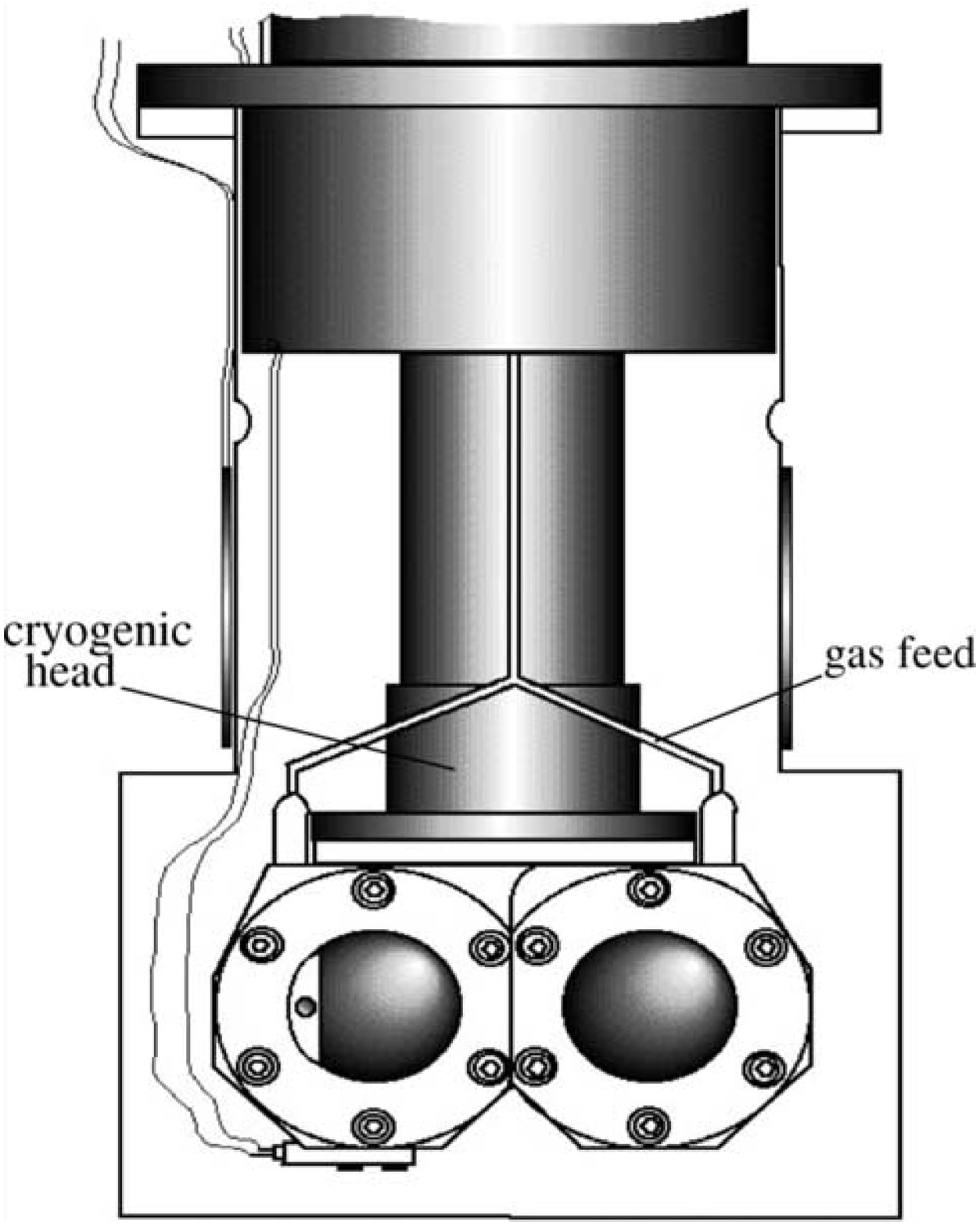}
\caption{Front view of the GANIL liquid target. The target is composed of two cells, 5 and 10 mm thick, respectively. Reprinted from~\cite{deVismes2002295}, with permission from Elsevier.}
\label{fig15b}
\end{center}
\end{figure}

A liquid hydrogen target composed of two cells (5 mm and 10 mm thick, respectively) has been developed at GANIL for reaction cross section measurements~\cite{deVismes2002295}. The target cells have a 20-mm diameter and are composed of 4.4 $\mu$m thick Harvar-foil windows (7 mg/cm$^{2}$ in total). Hydrogen gas is cooled down to a liquid phase at 20 K by the use of a cold head composed of two stages at 80 K and 20 K, respectively. A 1-mm hole on the side of the target (see Fig.~\ref{fig15b}) was used to measure the effective thickness of the target by energy loss difference. Due to atmospheric pressure operation, a significant curvature of the target profile occurs. The 10-mm thick target cell was measured to correspond to an average thickness of 11.2(1) mm. \\

\subsubsection{The PRESPEC liquid hydrogen target}
The PRESPEC collaboration aims at bridging the gap from the RISING collaboration at GSI and the upcoming HISPEC/DESPEC collaboration at FAIR. It aims at nuclear structure studies via gamma spectroscopy either from in-beam spectroscopy or from stopped beams. In this context, a liquid hydrogen target for relativistic-energy prompt gamma spectroscopy has been conceived at CEA Saclay.\\
The target is composed of a thermo-formed mylar cell (typically 200 $\mu$m thick) of few centimeters thickness and a large diameter of 75 mm. A large advantage of this geometry is to allow a free environment around the target, the cooling system being deported one meter upstream the target. The technique allows to build target cells from 10-mm to 80-mm. Typical 20-mm and 60-mm thick targets have been built (see Fig.~\ref{fig3}). It has been tested to support internal pressure over 14 bars, making the system extremely safe relative to potential over pressures.\\

\begin{figure*}
\begin{center}
\includegraphics[trim= 0mm 30mm 0mm 30mm,clip,width=14cm]{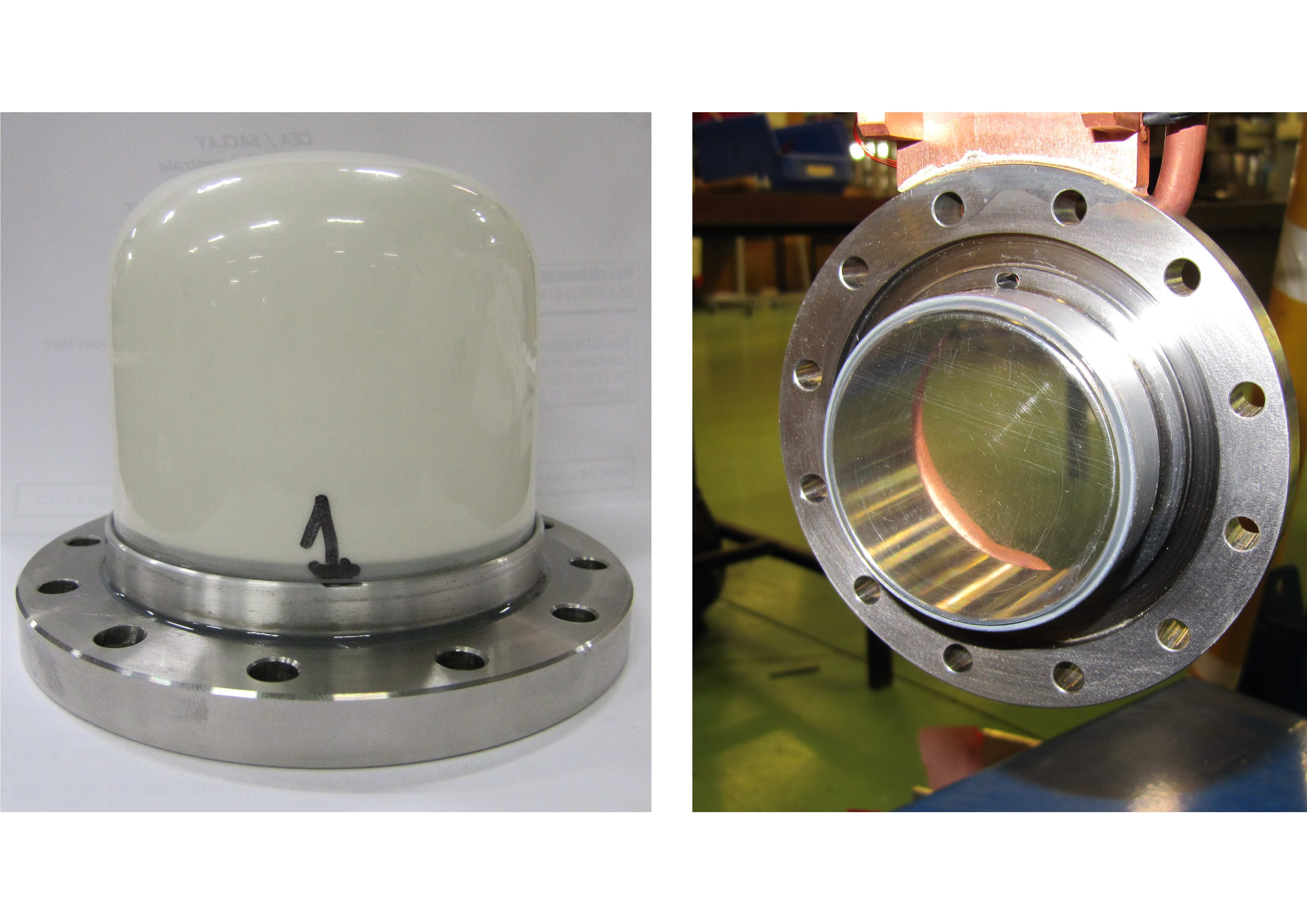}
\caption{(Color online) (Left) Photograph of a PRESPEC target cell with effective thickness of 61 mm. Targets from 10 to 80 mm can be built in one piece of 250-$\mu$m thick Mylar. (Right) Entrance window of the target (mounted to the hydrogen circuitry system) made of 100 $\mu$m-thick Mylar.}
\label{fig16}
\end{center}
\end{figure*}

Depending on the final target thickness, the Mylar thickness of the original foil is taken from 100 to 350 microns. The foil is warmed up to about 200 degrees and then pressed at the right depth. The process streches the foil and the final thickness of the exit window is usually about 40\% lower than the initial thickness of the foil. The target cell is then glued on a female flange (see left part of Fig.~\ref{fig16}). An entrance 100-micron thick window is processed in the same way and glued on the corresponding male flange (right part of Fig.~\ref{fig16}). \\

An in-beam test was performed at GSI in may 2011 to validate the complete system in view of upcoming experimental fast-beam campains. A target was succesfully cooled down and the inelastic scattering $^{54}$Cr(p,X)$^{54}$Cr$\gamma$ and inclusive knockout channels $^{54}$Cr(p,X)$^{53}$Cr$\gamma$ and  $^{54}$Cr(p,X)$^{53}$V$\gamma$ were measured with the PRESPEC setup composed of twelve Euroball Germanium clusters and the LYCCA recoil detection system~\cite{LYCCA}. Analysis is ongoing.\\

\subsubsection{The MINOS project: a H$_2$ target-TPC device for prompt $\gamma$ spectroscopy}
In order to reach the most neutron-rich nuclei produced at fragmentation facilities, one of the authors (AO) proposed to develop a new technique by using nucleon-removal from very exotic nuclei on a very thick liquid hydrogen target. The target is coupled to a proton tracker used to measure, on an event-by-event basis, the reaction-vertex position in the target. The project is called MINOS~\cite{minos}. By measuring the reaction vertex, one allows the use of targets of hundreds of millimeters with improved detection sensitivity, i.e. the Doppler correction is better than with a passive heavy-ion target. The only remaining limiting factor is that one has to ensure that the “second-interaction” probability in the target is low since the knockout fragment has to be identified after the target. For incident energies of E=200-400 MeV/nucleon, a typical length of 150 mm fullfils this condition. This development will induce a unique gain in detection sensitivity of more than an order of magnitude compared to experiments with solid heavy-ion targets.  In conjonction with the new generation AGATA array, the gain should reach a factor about several hundreds compared to existing setups, allowing the detailed spectroscopy of nuclei produced at less than 1 particle per second. \\

\begin{figure*}[t]
\begin{center}
\includegraphics[trim= 0mm 0mm 0mm 0mm,clip,width=12cm]{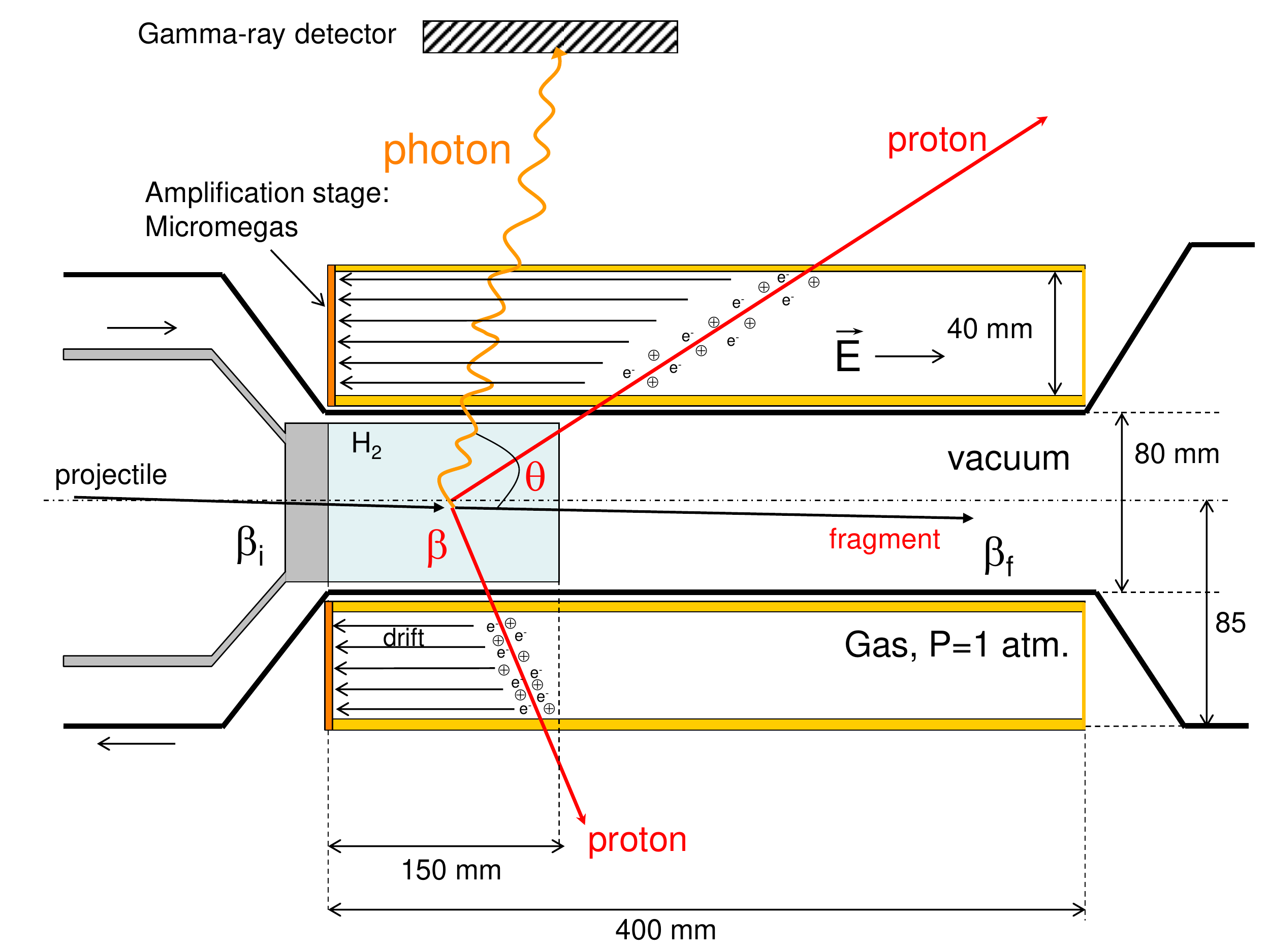}
\caption{Scheme of the MINOS target-TPC device, as imagined today. Quoted dimensions are preliminary. The final vertex position resolution is intended to be inferior than 3 mm FWHM.  The tracker allows to measure with precision the emission angle $\theta$ and the velocity $\beta$ of the fragment at the vertex position, essential for a good Doppler correction.}
\label{fig17}
\end{center}
\end{figure*}

A pure liquid H$_2$ target, used alone, has still some limitation due to the energy loss in the target and its thickness which, if larger than a centimeter, limits the angular resolution for the Doppler correction.  In case of a 150 mm thickness, the velocity spread is about 8-15\% for an initial velocity corresponding to 250-350 MeV/nucleon, depending on the considered incident ion. This example shows that such a thick target cannot be used for high-resolution gamma spectroscopy and the knowledge of the vertex position would then be needed.
 This is one of the reasons of the MINOSfor which a thick cryogenic LH$_2$ target is coupled to a light-particle tracker in order to reconstruct the vertex position. The development of such a device for intermediate-energy (p,Xp)-like reactions coupled to an hydrogen target is possible since (i) at least one light charged particle (proton) is emitted during the reaction, (ii) the energy loss of these protons in H$_2$ is most of the time small compared to their kinetic energy. The objective is to reach a 3-mm vertex-position resolution FWHM. In the case of a $(p,2p)$ measurement such as $^{63}$V(p,2p)$^{62}$Ti at 250 MeV/nucleon and a target thickness of 150 mm, the energy loss in the target represents 30\% of the total kinetic energy, i.e. a $\Delta \beta$=8\% velocity spread from $\beta$=0.61 to $\beta$=0.53 along the target thickness. A $\Delta$x=3 mm resolution in the vertex position gives a $\Delta \beta / \beta = $0.2\% velocity resolution and does not impact the angular resolution for the scattering angle needed in the Doppler-effect correction. An equivalent energy resolution is found in a 3-mm thick $^9$Be target, with approximately 20 times less atoms.cm$^{-2}$. Considering that deeply-bound proton removal reactions at intermediate energies have similar cross sections from H, $^9$Be or $^{12}$C, this leads to a net gain of a factor ~20 in statistics. \\

The target will have a similar design (a finger-shaped envelope) as the recent PRESPEC target development. The geometry and material of the target will be chosen to minimize the absorption at all angles (including 90 degrees in the laboratory frame). The angular straggling in hydrogen for such a thick target remains rather low. SRIM Monte-Carlo calculations for $^{110}$Zr at 350 MeV/nucleon give an angular straggling of approximately 3 mrad FWHM for a 150-mm thick target.\\
The trajectories of protons will be analyzed using a cylindrical time-projection chamber (TPC). The TPC will surround the target and be 300-400 mm long to allow the detection of charged hadrons emitted at small angles (see a schematic view in Fig.~\ref{fig17}) and in the momentum range from 300 to 1500 MeV/c. In the case of $(p,pn)$ reactions, it corresponds to the detection of recoil protons whereas in the case of knockout (p,2p) reactions both the recoil and the projectile-like protons will be detected in coincidence. Angular resolution should be of the order of 20 mrad or less.  Drift electrons will be amplified on the backplane of the TPC through a Micromegas amplification stage~\cite{Giomataris199629,Abgrall201125}. Charges will be induced on a pixelized plane with pads of the order of 2-4 mm$^2$. Signals will be digitized via a specific electronics and readout based on the GET (General Electronics for TPCs) developments~\cite{get}, a collaboration between CEA-IRFU, IN2P3 and MSU/NSCL.  

We performed simulations of the tracker with realistic events from Monte-Carlo simulations. Events have been generated with an intra-nuclear cascade code. The target radius is assumed to be 35 mm and the two tracker-cylinder radii are of 45 and 85 millimeters, respectively. The diameter is not fixed at this stage of the project. A smaller radius of 20 mm is under study. The following conclusions have been obtained from the two-systems $^{111}$Nb+p and $^{55}$Sc+p at 250 MeV/nucleon. Energy losses and detection thresholds have been considered.
The detection efficiency for $(p,2p)$ reactions is larger than 90\% with approximately 70\% of events with both protons detected. In case of $(p,3p)$ reactions, the detection efficiency is even higher since the exit channel contains more protons. Interestingly, $(p,pn)$ reactions are also detected with a significant efficiency of more than 70\%. \\

Simulations of the gamma detection system with the GEANT4 software have been performed for the same reactions to characterize the gain in sensitivity expected from this development. We considered the population of $^{54}$Ca(2$^+$) and $^{110}$Zr(2$^+$) via one-nucleon knockout. Both transitions have been considered as very short lived. We considered in each case (i) the foreseen development with a vertex position resolution of 3-mm FWHM and (ii) a "standard" $^9$Be target with a thickness limiting the velocity spread in the target to 5\% FWHM. The statistics corresponds to one week of beam time for a one-nucleon knockout cross section of 10 mb and a beam intensity of 1 particle per second. Simulations have been performed with the AGATA array with a 1$\pi$ configuration at forward angles located at 20 centimeters from the center of the target in each simulation. Simulations show that the figure of merit, taken as proportional to the ratio of the statistics to the energy resolution, reaches 15 in the case of $^{54}$Ca and 30 in the case of $^{110}$Zr. 

The lifetime of the populated states acts on the shape of the measured photopeaks. It was estimate through simulations. A distorsion of the lineshape is visible for lifetimes larger than 20 picoseconds (ps). For lifetimes from 20 ps to 100 ps, the shape of the measured photopeak can be used as a measurement of the lifetime with still a significant gain compared to the use of a thin target. In the case of a longer lifetime, the distorsion of the photopeak starts to be too large to maintain a good-enough sensitivity.

\subsection{Gaseous active targets}
For some specific studies, the detection of low energy recoil proton-like particles is necessary. Very thin targets are therefore necessary but obviously strongly affect the final statistics. One solution is the use of gaseous active targets.\\

At the end of the 1960s, the first position sensitive gaseous detectors were developped, following the invention of multiwire chambers.  These first detectors were used for particle physics experiments and large evolutions have been made such as micro-strip gas chambers, GEMs (Gas Electron Multiplier), etc... Very soon it appeared that the spatial resolution could be largely improved by measuring the induced charge on pads or strips in the vicinity of amplifying wires or surfaces. First detectors specifically dedicated to nuclear physics have been developed in that sense~\cite{Christie1987466,Kimura1990190}. Two TPCs (Time Projection Chambers) have been developed in the 80s-90s with a specific design for reaction studies with stable beams: the IKAR chamber~\cite{Dobrovolsky19831,Vorobyov1988419} and the MSTPC (Multi Sample Tracking Proportional Chamber)~\cite{Mizoi1999112}. In the following, we present recent developments for active targets dedicated to inverse kinematics reaction with exotic beams.\\

\subsubsection{The GANIL active target MAYA}
\begin{figure}[!]
\begin{center}
\includegraphics[trim= 0mm 0mm 0mm 0mm,clip,width=7.5cm]{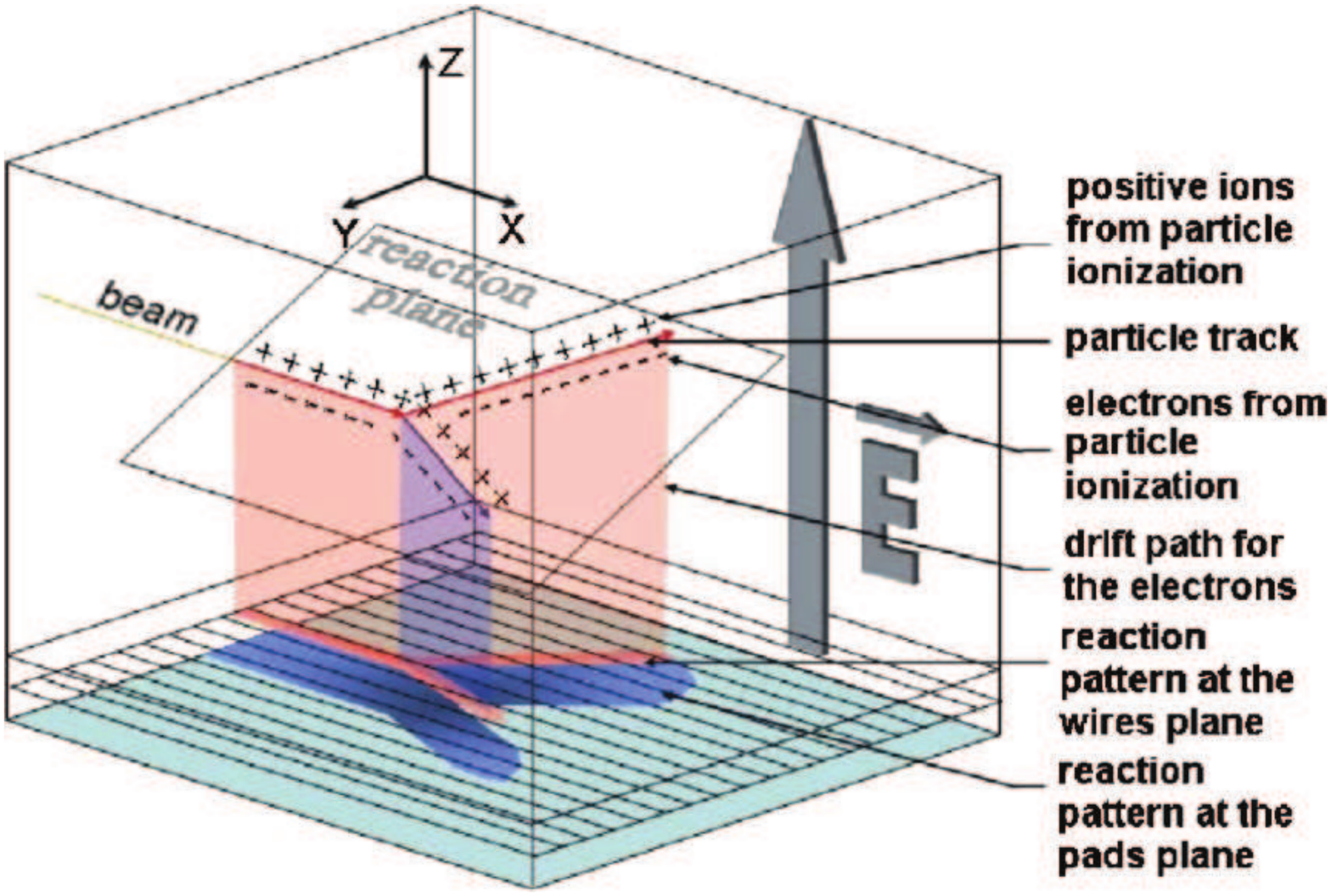}
\includegraphics[trim= 0mm 0mm 0mm 0mm,clip,width=7.5cm]{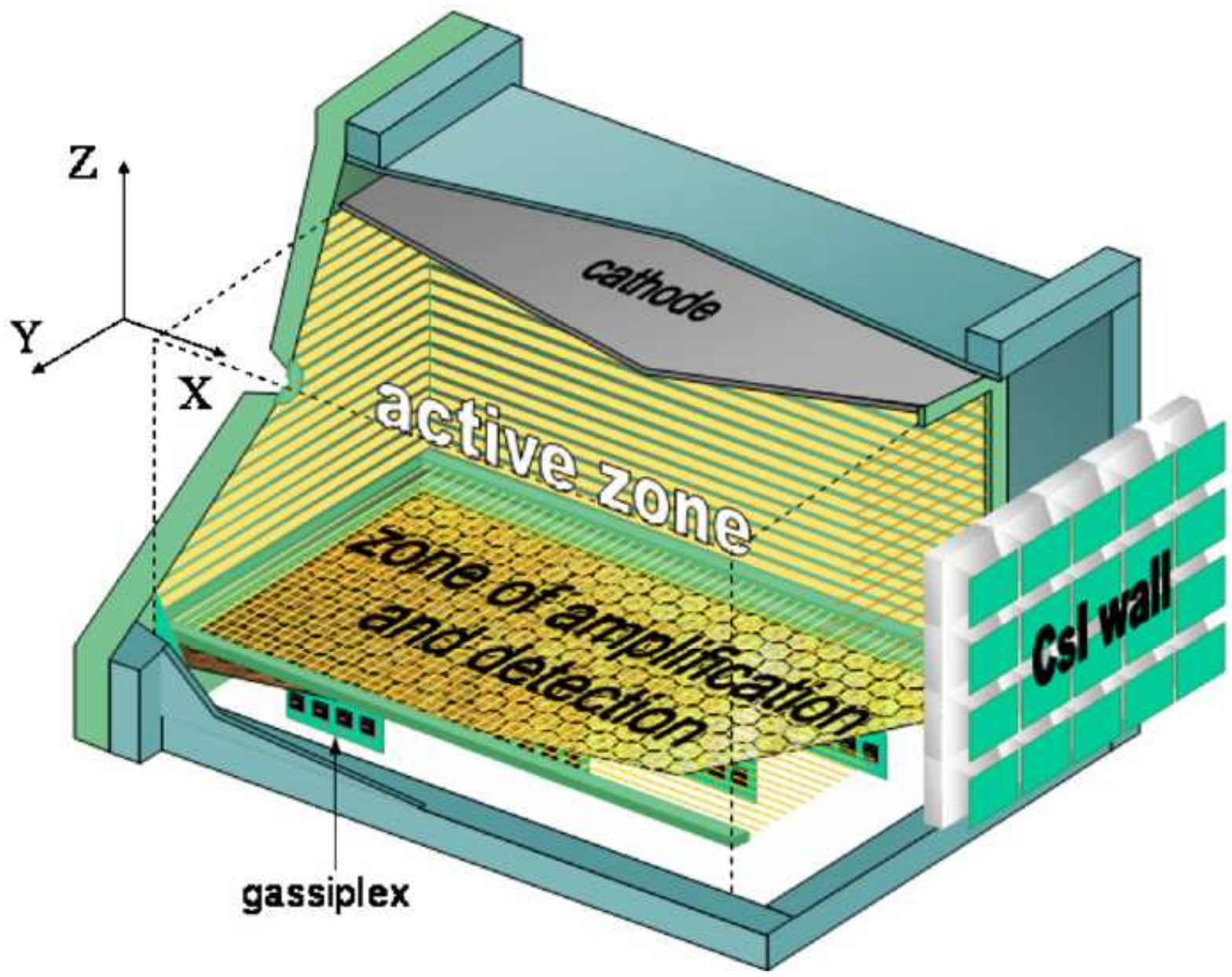}
\caption{(Top) Principle of a Time Projection Chamber. (Bottom) Scheme of the MAYA detector. Reprinted from~\cite{Demonchy2007341}, with permission from Elsevier.}
\label{fig18_19}
\end{center}
\end{figure}

The MAYA detector has been developed at GANIL by W. Mittig and his collaborators~\cite{Demonchy2007341}.\\
MAYA works essentially as a drift chamber where the filling gas also serves
as the target. Two main zones can be identified within the detector: the conversion and drift
zone where the reaction takes place, and the amplification zone where the signal readout is made. The volume of the active zone is 28$\times$26$\times$20 cm$^3$, defined by a cathode
plane at the top and the amplification zone at the bottom. The amplification zone
consists of a Frisch grid in the upper part, an anode wire plane below that, and
a segmented cathode in the bottom part. The stainless-steel
container of the detector, with a 1-centimeter diameter Mylar window, was tested for gas
pressures up to 3 atm incuding explosive gases. As an example, a 6-micron thick Mylar window was used for a 2 atm pressure. Figure~\ref{fig18_19} shows a schematic view of the detector.
The anode wires in the amplification zone are parallel to the beam axis.  
The anode wire plane is just above a lower cathode, which is segmented into 35×35
hexagonal pads, each of them has a 5 mm side. The cathode plane is connected to
a set of Gassiplex chips. The signals induced in the pads are recorded and
stored in the Gassiplex through a Track-and-Hold procedure, triggered by a signal
from the wires, until they are sent to data acquisition.
In a typical event the beam enters MAYA through the Mylar window after
going through several monitor detectors and ionizes the filling gas. If a projectile hitsa gas atom, a reaction takes place. The electrons
from the atoms in the gas that have been ionized by the reaction products
drift to the amplification zone by means of an electric field applied between the
upper cathode and the proportional wires, while the Frisch grid is kept grounded.
The induced charges in each cathode pad form a projected image of the particles
trajectories. The homogeneity of the electric field is maintained by metallic strips
covering the sides of the detector, except at the back, where they are replaced by
field wires to reduce interaction with the forward escaping particles. The electric
field can be set as high as 15 kV in the upper cathode, and 5 kV in the proportional
wires, depending mainly on the pressure and the detection energy threshold of the
specific particles.
Inverse-kinematics reactions generate scattered particles in a large energy domain. High
energy light particles cannot be stopped in a reasonable gas volume and pressure, and
escape from the active volume. At forward angles, the escaping particles are stopped and identified in
twenty Cesium-Iodide (CsI) crystal detectors, arranged in five columns
and four rows which cover the exit face of the detector.
Outgoing light pulses were used to identify the mass and
charge of the stopped particle.\\

MAYA has been used with several gases to perform transfer or inelastic scattering experiments. Some of these experiments were performed with an hydrogen gas:
\begin{itemize}

\item $^{11}$Li(p,t)$^9$Li at 3 MeV/nucleon~\cite{PhysRevLett.100.192502}. The correlation of the two neutrons in the halo
of $^{11}$Li halo nucleus was studied  at TRIUMPH. All the outgoing particles were detected in the
C$_4$H$_{10}$ gas at about 100 mbar pressure and, in some cases, stopped in solid-state
detectors placed at forward angles.

\item $^{56}$Ni(d,d')$^{56}$Ni at 50 A MeV~\cite{PhysRevLett.100.042501}. The inelastic scattering on deuterons in inverse
kinematics was used to populate the giant monopole resonance in $^{56}$Ni at GANIL.
The position of the resonance is related to the nuclear matter incompressibility.
MAYA was operated with deuterium at a pressure of 1050 mbar.
\end{itemize}

\subsubsection{TPC projects for nuclear-structure studies with radioactive beams}
ACTAR is a new active target/time-projection chamber project carried by European laboratories. ACTAR incorporates recent developments in gas detector technology and a
newly-designed electronic system. The instrument
will be able to take advantage of the exotic beams produced at
in-flight fragmentation facilities, as well as the high-quality ISOL beams which will
become available at SPIRAL2. Hydrogen-induced reactions in inverse kinematics represent a large part of the physics motivations of ACTAR. A new TPC project called AT-TPC, initalized by W. Mittig, is being completed at the NSCL. Finally, a collaboration between NSCL and RIKEN is developping another large TPC for nuclear-matter equation-of-state studies.\\

\subsubsection{Gas-jet targets for storage rings}
The detection of very forward angles in the center of mass in (p,p') inelastic scattering experiments at relativistic energies may require very thin targets. The use of storage rings allows the use of target thicknesses of about 10$^{14}$ cm$^{-2}$ hydrogen atoms and still keep a sufficient luminosity~\cite{Amadio2008191}.

\begin{center}
\begin{table*}
\begin{tabular}{|c|c|c|c|c|c|c|c|c|}
\hline
Phase&Name&Host lab.&Ref.&thickness&diameter&T&Windows / thickness&cooling\\
&&&&(mm)&(mm)&(K)& / ($\mu$m)&(min)\\
\hline
Liquid&CRYPTA&RIKEN&\cite{Ryuto20051}&10 -- 50&40&20&Havar / 12&100\\
&GANIL liquid&GANIL&\cite{deVismes2002295}&5,10&20&20&Harvar / 8.8&60\\
&PRESPEC&CEA Saclay&-&20-80&70&20&Mylar / 300&500\\
&MINOS$^{\star}$&CEA Saclay&\cite{minos}&80-200&[40-70]&20&Mylar / 200&500\\
Solid&RIKEN solid&RIKEN&\cite{Ishimoto2002304,matsuda2011}&1-5&30&3&none&\\
&GANIL&GANIL solid&\cite{Dolegieviez200632}&1-5&10&4&Mylar / 24&240\\
&CHyMENE$^{\star}$&CEA Saclay&\cite{gillibert}&0.05-5&10&17&none &120\\
&RIKEN ultra-thin$^{\star}$&RIKEN&-&0.12&unknown&unknown& / 0.025 -- 0.2&unknown\\
Gas&MAYA&GANIL&\cite{Demonchy2007341}&300, P$ <$3 bars&10 mm&ambiant& Mylar / 0.9 -- 6&-\\
\hline
\end{tabular}
\caption{Main technical characteristics of non-polarized H$_2$ or D$_2$ targets described in this review. Targets marked with a $\star$ are projects in development.}
\end{table*}
\end{center}

\section{Polarized targets}
\label{section2}
Since its first application to nuclear and high-energy physics experiments~\cite{Chamberlain63}, polarized proton targets have been mostly used with high-energy electron or proton beams. Developments of polarized proton targets~\cite{Goertz02,Crabb97} have been made to optimize the target performance to fit with the experiments. On the other hand, requirements in radioactive nuclear beam experiments are largely different~\cite{RMF2007,Uesaka04}.

RI beams are produced by nuclear reactions and their typical intensity is less than 10$^{6}$ per second. Therefore, in order to perform reaction studies with low-intensity secondary beams, the target should be thick. If we set a criteria of luminosity at ${\cal L} > 10^{23}$ /cm$^2$/s, a gas jet target with a thickness of $10^{14}$/cm$^2$ does not fit the requirements, except in the case of storage ring experiments. Indeed spin-polarized solid targets are needed in fixed-target experiments.

Any spin-polarized solid target is a compound of hydrogen and other elements. Thus it is necessary to identify reaction events from hydrogen from those due to other atoms. Detection of recoil protons facilitates the event identification. The recoil protons scattered at forward angles in the center-of-mass system have an energy as low as 5--10 MeV. 
External magnetic field which is indispensable in polarized targets should be low enough to enable the low-energy proton detection. 

Low magnetic field operation is needed even when the recoiled particles have sufficiently large energy but the trajectories have to be determined with a good angular resolution on the order of 10$^{-3}$~radian. This is the case in 
$(p,pN)$ experiments in inverse kinematics condition. 
 
 Traditional dynamic-nuclear-polarization (DNP) targets~\cite{Goertz02,Crabb97}
 use thermal polarization of electrons to polarize protons and thus a magnetic field higher than a few Tesla is required together with a sub-Kelvin temperature. For this reason, one would find it difficult to apply a DNP target to RI-beam experiments as it is. 

There are two possible ways to enable the low-magnetic field operation: one is a spin-frozen operation of traditional DNP targets and the other is a proton polarization based on an electron polarization in a photo-excited triplet state of aromatic molecule. The possibility of spin-frozen target is pursued by the ORNL (Oak Ridge National Laboratory) - PSI (Paul Scherrer Institute) collaboration.
The CNS-RIKEN groups has constructed a polarized proton solid target based on triplet-state polarization and applied it to radioactive nuclear beam experiments at RIPS, RIKEN.

\subsection{CNS-RIKEN polarized target}
Proton polarization  in aromatic molecules is achieved by transferring the large population difference among Zeeman sub-levels in the photo-excited triplet state of aromatic molecule~\cite{Kesteren85}. Here, the value of the population difference depends neither on its temperature nor on strength of external magnetic field, which makes it possible to polarize protons up to several tens of percent \cite{Iinuma00,Wakui05,Takeda02} in a modest condition of temperature higher than 77 K and a magnetic field of lower than 0.3~T. 

\begin{figure}[htbp]
 \centering
 \resizebox{8cm}{!}{\includegraphics{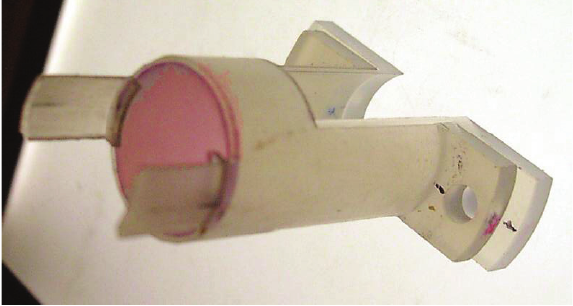}}
  \caption{(Color online) A naphthalene target mounted on a holder. \label{fig20}}
\end{figure}

The target material is a single crystal of naphthalene (C$_{10}$H$_{8}$) doped with a small amount (0.005 mol\%) of pentacene. The target is formed by the Bridgman crystallisation after zone melting refinement. 
For use in a scattering experiment, the crystal is shaped as a disk with a thickness of 1 mm. The cross section of the target is as large as 14 mm in diameter to cover a spot size of RI-beams.
Fig.~\ref{fig20} shows a picture of the naphthalene target mounted on a holder. The purple color originates from pentacene molecules (C$_{22}$H$_{14}$). The target holder is made of hydrogen-free polychlorotrifluoroethylene to avoid background production.

\begin{figure*}[htbp]
 \centering
 \resizebox{15cm}{!}{\includegraphics{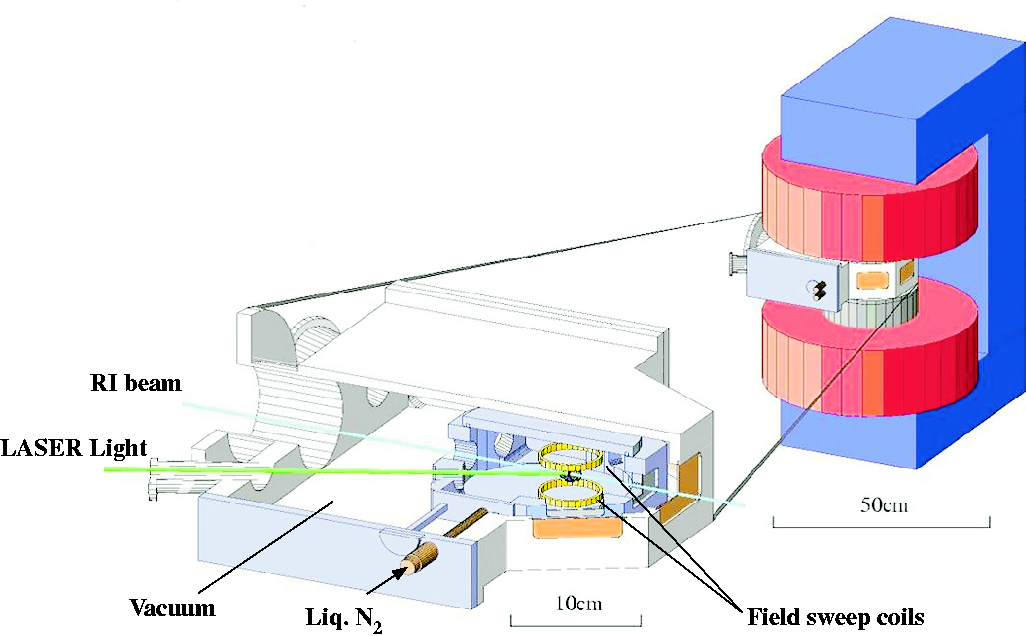}}
  \caption{(Color online) The CNS-RIKEN polarized proton target system.  \label{fig21}}
\end{figure*}

A scattering chamber supporting the target is put in a magnetic field of about 0.1~T produced by a C-type magnet with a pole gap of 100~mm and a pole diameter of 220~mm. A magnetic field homogeneity of $\Delta B\sim$ 4 mT is achieved at $B=$0.3~T over the target volume.  The scattering chamber shown in Fig.~\ref{fig21} has a twofold structure: a small cooling chamber including the target is placed inside an outer scattering chamber. The cooling chamber is connected to the scattering chamber with a stainless rectangular channel and pipes to feed  the cooling chamber with cold nitrogen gas. The volume between the two chambers is pumped to vacuum to reduce thermal contact between the chambers and to reduce materials which cause background events. Temperature in the cooling chamber is monitored continuously with a platinum thermometer and is kept at 100~K by adjusting the flow rate of cold nitrogen gas.\\

Each chamber has six windows, two for laser light and four for ions. Laser light from Ar$^+$-ion lasers
~\cite{Wakui05} is brought to the target crystal through two glass windows placed at 30 degrees and used to populate photo-excited triplet states of pentacene molecules. Heavy ion beams are injected through
an entrance window made of 6-$\mu$m thick Havar foil, while heavy ions scattered at forward angles and protons scattered at backward angles go through windows made of 12-$\mu$m thick Kapton films.\\

Field sweeping for polarization transfer from electrons to protons is done with a pair of coils around the target. The sweeping range and rate are 3.9 mT and 0.43 mT/s, respectively.\\

Electron polarization produced by laser irradiation is transferred to protons with a cross polarization method~\cite{Henstra90}. In the method, a microwave at the electron spin resonance frequency is irradiating the target while an external magnetic field is varied simultaneously to cover the inhomogeneously broadened linewidth. To keep a high penetrability of recoiling protons, a thin copper film loop-gap resonator (LGR)\cite{Ghim96} is used as the microwave resonator. The copper film LGR is a thin Teflon cylinder with copper platings on both sides. The resonant frequency is determined by plating geometry, the thickness of Teflon, and the length of the resonator. The thicknesses of Teflon sheet and copper plating are 25~$\mu$m and 4.4~$\mu$m, respectively. The actual resonant frequency is 2--3 GHz depending on the design.
\begin{figure}[htbp]
 \centering
  \resizebox{8cm}{!}{\includegraphics{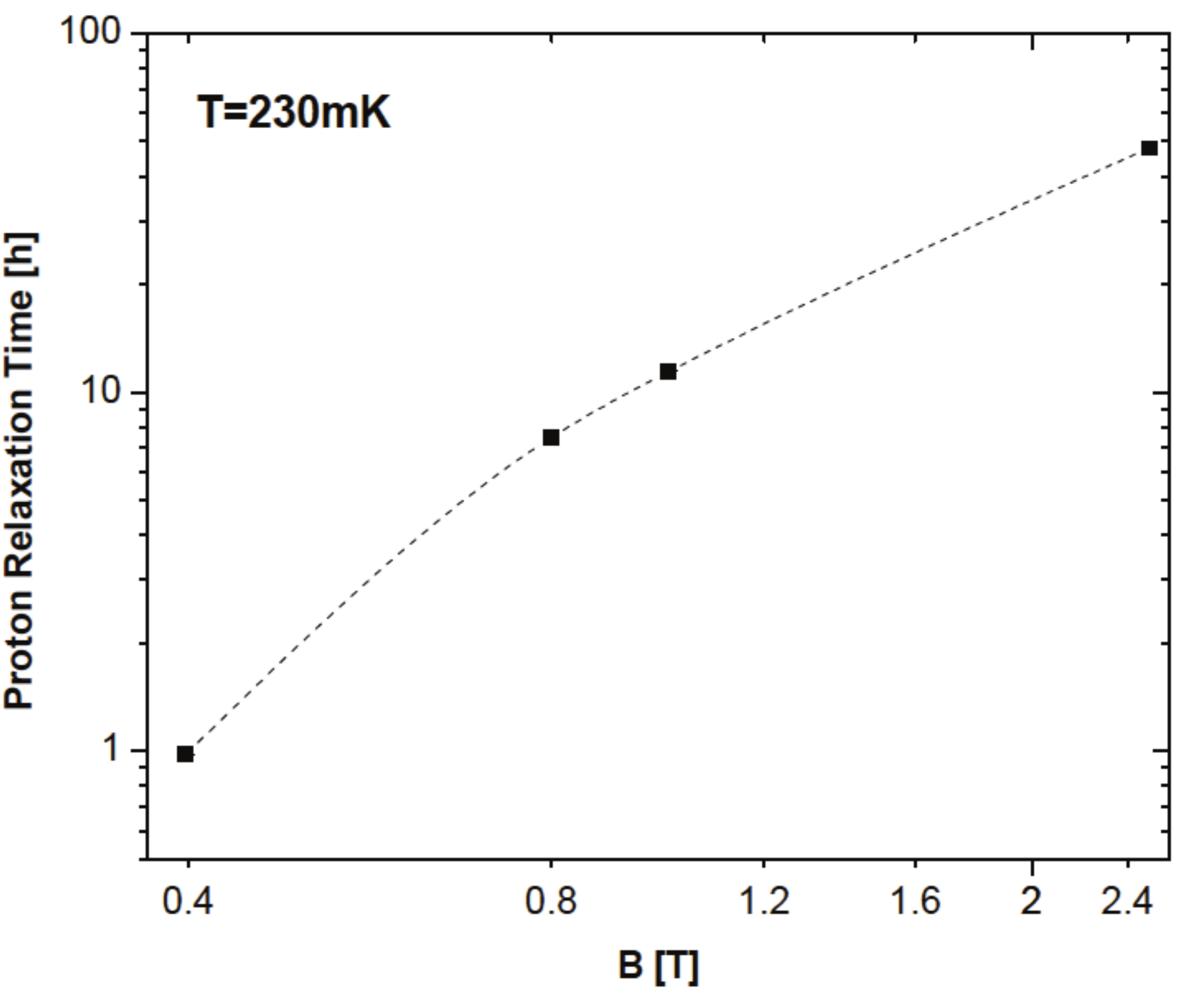}}
  \caption{Magnetic field dependence of the spin relaxation time at 230~mK. \label{fig25}}
\end{figure}
The LGR of the current design has a large diameter (16~mm) to length (20~mm) ratio of 0.8 to ensure sufficient acceptances both for incident/scattered heavy ions and for laser light. This large ratio, however, causes unwanted coupling of the resonator to surrounding metal materials such as NMR coils, cables, and so on. The couplings cause a decrease of a microwave strength at the target position.To avoid those unwanted couplings, a shield made of 12~$\mu$m aluminum foil was placed around the target. The proton polarization during the experiments was monitored by a pulse NMR method at 2.94 MHz. The NMR amplitude as a function of time is shown in Fig.~\ref{fig25}. \\

\begin{figure}[htbp]
 \centering
  \resizebox{8cm}{!}{\includegraphics{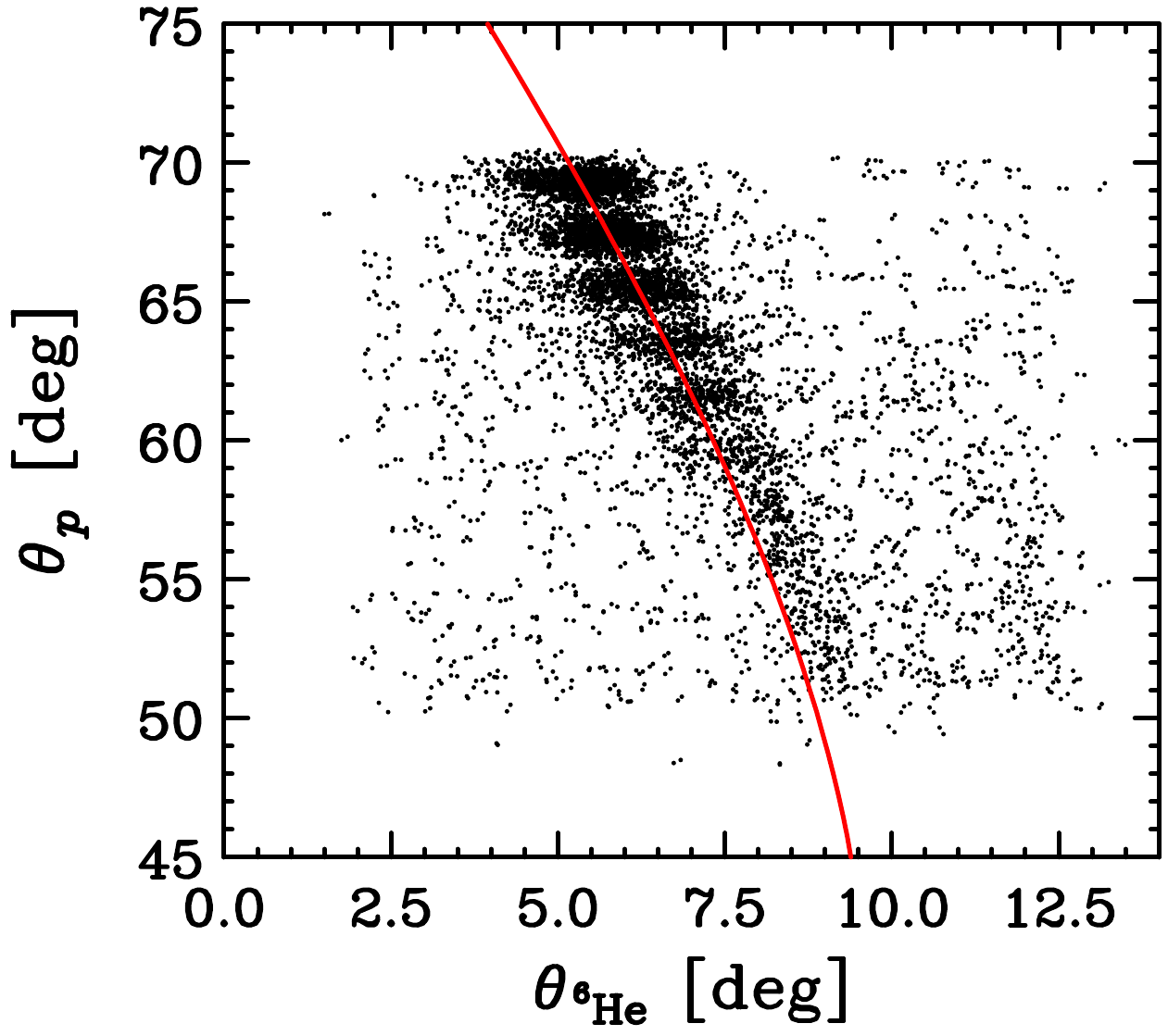}}
  \caption{Kinematical correlation of protons and ${\rm ^6He}$. Dots represent experimental data, while the solid line a result of relativistic kinematics calculation. \label{fig22}}
\end{figure}

The polarized target technique was applied to a RI-beam experiment by using the projectile fragment separator RIPS~\cite{Kubo1992} at RIKEN. In the experiment, a ${\rm ^6He}$ beam at an energy of 71 MeV/nucleon and an intensity of 2$\times 10^5$ pps bombarded the polarized target. The maximum value of the proton polarization was found to be 20.4~\%. Although a decrease of the proton polarization was observed during the ${\rm ^6He}$-beam irradiation, the gradient was as small as 0.1~\%/hour and the polarization was persistent for several days. The low magnetic field operation made it possible to detect the low-energy recoiling protons and to identify the reaction events. Figure~\ref{fig22} exhibits the correlation of scattering angles of proton and ${\rm ^6He}$, and a clear locus corresponding to the $p$-${\rm ^6He}$ elastic scattering is seen. As a result, data of a vector analyzing power, together with cross section, were successfully measured for the elastic scattering of polarized proton and ${\rm ^6He}$  in the angular range of 39--78$^{\circ}$ in the center of mass system~\cite{PhysRevC.82.021602,Sakaguchi11}. \\

\subsection{ORNL-PSI polarized target}
  State-of-the-art DNP targets can be, in principle, applied to RI-beam experiments by introducing "spin-frozen" operation. A collaboration of the ORNL and the PSI is pursuing this possibility~\cite{UrregoBlanco07,UrregoPhD}. 
  
The target material is a polystyrene plastic (${\rm C_{8}H_{8}}$) with a free nitroxyl radical TEMPO~\cite{Rozantsev70,Bunyatova04}. Plastic is a flexible material and can be formed to any shape to fit experimental requirements. Availability of a thin target with a thickness of less than 1~mg/cm$^2$ is a big advantage in applications to experiments with low-energy RI-beams. The target film is attached to a copper frame which is in thermal contact with the mixing chamber of a dilution refrigerator. Temperature of the target is monitored by a ruthenium oxide thermometer placed on the copper frame.

\begin{figure}[htbp]
 \centering
  \resizebox{8cm}{!}{\includegraphics{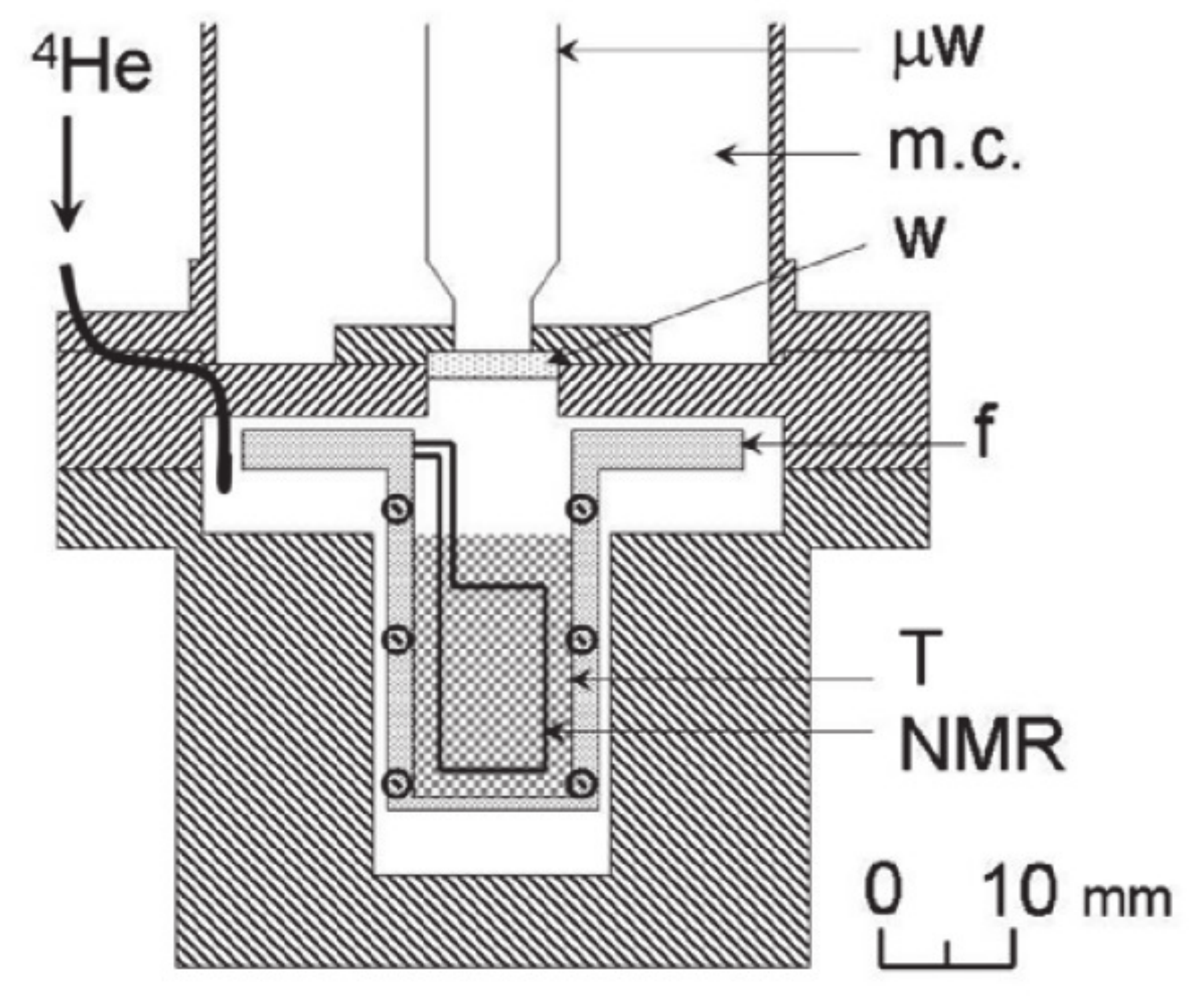}}
  \caption{A schematic figure of ORNL-PSI polarized target. \label{fig23}}
\end{figure}

A cryogenic system composed of the ${\rm ^3He}$-${\rm ^4He}$ dilution refrigerator and a ${\rm ^4He}$ cryostat is used to cool the copper frame and superconducting coils, and to prepare superfluid liquid helium~\cite{vandenBrandt90}. The superfluid liquid helium is directly supplied to surfaces of the target film and keeps the target cooled.\\

In the polarization mode, a high magnetic field of 2.5~T is added on the target and an electron Boltzmann polarization of about 100\% is produced at 0.2~K. Irradiation of microwaves at 70~GHz with a modulation of 1~kHz leads to a transfer of the electron polarization to protons. Proton polarization of 30\% has been achieved~\cite{UrregoBlanco07} when no beam is injected to the target.
\begin{figure}[htbp]
 \centering
\includegraphics[trim= 0mm 0mm 0mm 100mm,clip,width=8cm]{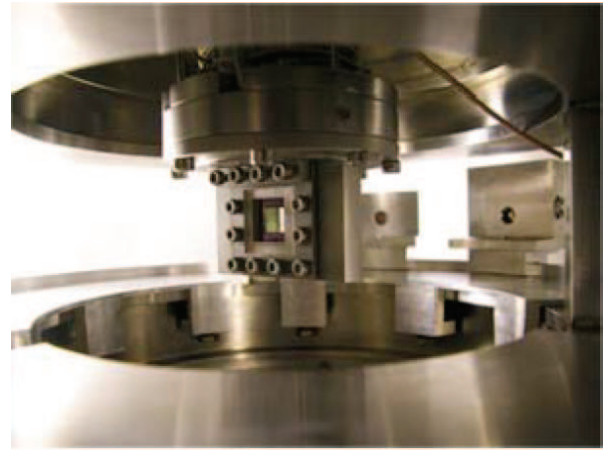}
  \caption{A photo of ORNL-PSI polarized target \label{fig24}}
\end{figure}
In the frozen spin mode, the microwave irradiation is interrupted and the magnetic field strength is reduced. The proton polarization survives within its spin-relaxation time. The spin-relaxation time 
decreases as the temperature increases and the magnetic field decreases. Thus it is essentially important to keep the target at sufficiently low temperature, typically much lower than 0.3~K. Removal of heat load from the energy loss of the heavy ions is another critical point.  

A beam irradiation test of the polarized target was conducted with a 38-MeV ${\rm ^{12}C}$ beam at the PSI. The polarized target was operated at 0.8~T to avoid the rapid spin relaxation at lower magnetic field. Peaks originating from resonance states in ${\rm ^{13}N}$ at $E_{x}=3.50$ and 3.55~MeV were clearly observed~\cite{UrregoPhD}. A rapid decrease of the proton polarization due to irradiation of the ${\rm ^{12}C}$ beam has also been observed.

\subsection{Discussions for future polarized proton targets}
As shown in the last two sections, challenges to build polarized proton targets for RI-beam experiments have been just initiated. There is much room for future improvement. In this section, direction of future development is discussed.

The characteristics of the two targets are compared in Table~\ref{tbl:comparison}.  Use of organic material carries an advantage, over lithium hydride and ammonia used in high-energy experiments, that one can easily prepare thin targets needed in low and intermediate energy heavy-ion experiments. Although the materials have a small dilution factor, which is defined as a ratio of polarizable nucleons to the whole nucleons, it does not cause a serious problem because the reaction events from hydrogens can be identified by detecting low-energy protons.

\begin{table}[htbp]
 \begin{center}
 \caption{Comparisons of the two targets.\label{tbl:comparison}}
 \begin{tabular}{lcc}
 \hline\hline
                   &   CNS-RIKEN & ORNL-PSI  \\
 \hline
     Material &     Naphthalene  & Polystyrene \\     
&   (${\rm C_{10}H_{8}}$) &  (${\rm C_{8}H_{8}}$) \\     
     Dilution factor & 0.063 & 0.077 \\
     Thickness (typical) & 100~mg/cm$^2$ & 0.1--3~mg/cm$^2$ \\ 
Diameter&14 mm&16 mm\\
     Temperature & 100~K & 0.3~K \\
     Magnetic field & 0.1--0.3~T & 0.4--0.8~T \\
     Polarization (w/o beam)  & 20\% & 30 \% \\ 
     Polarization (w/ beam) & 15\% &  13 \% \\
 \hline\hline
 \end{tabular}
\end{center}
\end{table}

The largest merit of the ORNL-PSI target is that there is almost no limit in its thickness: a thinner target is highly preferable in low-energy experiments and a thicker target helps to improve luminosity at higher energies. On the other hand the CNS-RIKEN target has less flexibility in the thickness for two reasons. One is that the present material, naphthalene is too fragile to prepare a target thinner than 10~mg/cm$^2$. The other is the requirement for  laser power. Since the required laser power increases almost linearly with the target thickness, dozens of Ar$^{+}$-ion laser is needed to polarize 1~g/cm$^{2}$ target. Future reinforcement of light sources, most probably through development of new lasers, is highly desired.

\begin{figure*}[t]
\begin{center}
  \includegraphics[trim= 0mm 0mm 50mm 130mm,clip,width=12cm]{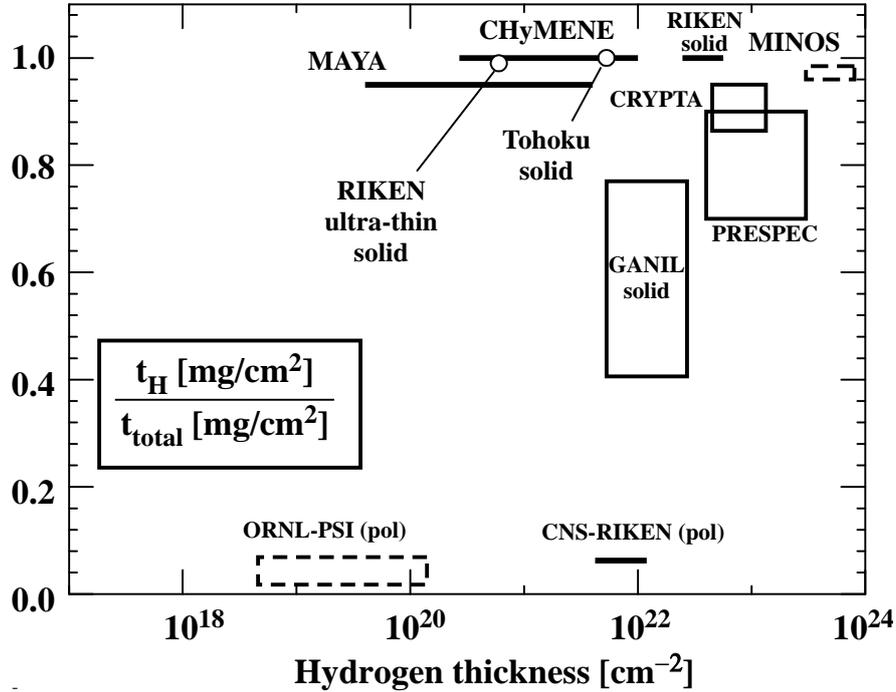}
  \caption{ Comparison of the thickness and purity of targets discussed in the present review. Exisitng targets are indicated with continuous lines or circles whereas dashed symbols are used for targets still in R\&D phase.\label{fig26}}
\end{center}
\end{figure*}

Values of achieved polarization before beam irradiation are comparable in both cases and are a factor of 2--3 lower than the goal values of 70--80\%.
In both cases loss of polarization were observed after irradiation of charged-particle beams.
It should be noted, however, that the reasons are completely different. 
In the CNS-RIKEN target, decrease of the polarization occurred in a gradual manner: the proton polarization in the ${\rm ^8He}$-beam experiment in 2007 was found to decrease from 15\%  to 11.3\% after an irradiation of 5$\times 10^{10}$ ${\rm ^8He}$ particles (in 6 days) at 71~MeV/nucleon. The spin-relaxation time before and after the beam irradiation were 0.077~hour${^{-1}}$ and 0.154~hour$^{-1}$, respectively.
Production of paramagnetic centers is considered to be responsible for the depolarization. 
The proton polarization in the ORNL-PSI setup was found to decrease more rapidly, from $\sim$18\% to $\sim$8.5\% in 24~min under an irradiation of 2$\times 10^{6}$ ${\rm ^{12}C}$ particles per second (3$\times 10^{9}$ particles in total).  A temperature increase of 55~mK is responsible for the increase of relaxation rate. After interruption of the beam irradiation, the relaxation time recovered to the initial value. This means that production of paramagnetic centers is not the problem in the ORNL-PSI setup\\

In conclusion, the future development of polarized targets should include (i) at CNS-RIKEN, the reinforcement of light source to improve the polarization. New target material with capabilities superior to naphthalene should also be searched for. (ii) At ORNL-PSI, further improvement of heat-removal efficiency should be pursued. (iii) Finally, a polarized proton gas-jet target in EXL setup of FAIR would be a promising approach.

\section{Summary}
As detailed in this article, a variety of activities has been initiated to develop hydrogen targets for RI-beam  experiments. It should  be emphasized that this diversity is inevitable because required performances are diﬀerent depending on applications: for instance, some experiments may need very thin targets for low-energy recoil particles or require a minimum of material around the target for a 4$\pi$ detection of gamma-rays. To illustrate the wide range of use covered by all the above-mentioned targets, the target thickness (cm$^{-2}$) and its purity (taken as the ratio of hydrogen to all the materials in mg/cm$^{2}$) are plotted in Fig. ~\ref{fig26}. 
The demand is less heavy in inclusive $\gamma$  spectroscopy experiments at intermediate incident energies such as those encountered at fragmentation facilities  like RIKEN or GSI. A high luminosity, a low background condition and minimized absorption around the target are most important.  Pure liquid or solid targets are favorable in this respect and, actually, the RIKEN liquid hydrogen target has produced many successful results. Other liquid hydrogen targets are developped at Ursinus university, USA, or at CEA Saclay, France. The quest to achieve highest possible luminosity, keeping the Doppler correction capability as much as possible, leads the CEA Saclay to combine the liquid hydrogen target and surrounding tracking detectors with the MINOS project. \\
In  proton-induced  knockout  reactions, purity  and suppression of multiple scattering eﬀects are important. Both can be carried out with solid windowless targets by the Japanese group from Tohoku university, Japan.  Recently, the technique has  been sophisticated by the introduction of para-hydrogen.\\
 Low-energy  experiments  require  targets  as  thin  as 1 mg/cm$^2$ or even thinner.  This is also the case in missing mass spectroscopy experiments, such as $(p, p')$ reaction studies at forward angles in the center-of-mass system.  Time-projection chambers (TPC), such as the  active-target MAYA from GANIL, France, are one of the solutions and have been proven to be powerful in specific cases. Several new-generation TPC projects are currently ongoing worldwilde, partly motivated by the use of hydrogen-induced direct reactions. Other solutions based on pure solid H$_2$ targets are being  proposed:   hydrogen film  extrusion  with  CHyMENE  or  nano-membrane in RIKEN ultra-thin solid target.  Considerable progress in this ﬁeld is foreseen for the next years.\\
Attempts  to  provide  polarized  proton  targets  have been just initiated by the CNS-RIKEN and the ORNL-PSI groups.  Improvement in degree of polarization is most urgent in the developments.  With a factor of 2$-$3 improvements, the polarized target will become a standard tool in RI-beam experiments.\\

Part of this work is supported by the European Research Council through the Grant 258567-MINOS, the FP7 Idea program, and the Grant-in-Aid No. 17684005 of the Ministry of Education,
Culture, Sports, Science, and Technology of Japan. The authors would like to thank N. Aoi (RCNP, Japan), P. Dolegieiviez (GANIL, France), Z. Dombradi, D. Solher (ATOMKI, Hungary), A. Corsi, J.-M. Gheller, A. Gillibert, C. Louchart (CEA Saclay, France), Y. Matsuda (Tohoku University, Japan), M. Nishimura (RIKEN, Japan) for their essential help during the preparation of the manuscript.

\bibliography{h2review}
\bibliographystyle{ieeetr}

\end{document}